\documentclass[preprint]{aastex}
\usepackage{hyperref}
\usepackage{graphicx}
\usepackage{epstopdf}
\usepackage{lineno}

\newcommand{\Fermi}{\textit{Fermi}}
\newcommand{\Swift}{\textit{Swift}}
\newcommand{\INTEGRAL}{\textit{INTEGRAL}}
\newcommand{\Suzaku}{\textit{Suzaku}}
\newcommand{\AGILE}{\textit{AGILE}}
\newcommand{\MAXI}{\textit{MAXI}}
\newcommand{\rb}[1]{\raisebox{1.5ex}[-1.5ex]{#1}}

\shorttitle{\Fermi\ GBM Gamma-Ray Burst Catalog}
\shortauthors{von Kienlin et al.}

\begin{document}

\title{The 2$^{\rm nd}$ \Fermi\ GBM Gamma-Ray Burst Catalog: \\
  The First Four Years}

\author{
Andreas~von~Kienlin\altaffilmark{1},
Charles A.~Meegan\altaffilmark{2},
William S.~Paciesas\altaffilmark{3},
P.~N.~Bhat\altaffilmark{2},
Elisabetta~Bissaldi\altaffilmark{4,5},
Michael~S.~Briggs\altaffilmark{2},
J.~Michael~Burgess\altaffilmark{2},
David~Byrne\altaffilmark{6},
Vandiver~Chaplin\altaffilmark{2},
William~Cleveland\altaffilmark{3},
Valerie~Connaughton\altaffilmark{2},
Andrew~C.~Collazzi\altaffilmark{7,8},
Gerard~Fitzpatrick\altaffilmark{6},
Suzanne~Foley\altaffilmark{6},
Melissa~Gibby\altaffilmark{9},
Misty~Giles\altaffilmark{9},
Adam~Goldstein\altaffilmark{2},
Jochen~Greiner\altaffilmark{1},
David~Gruber\altaffilmark{1,11},
Sylvain~Guiriec\altaffilmark{10},
Alexander~J.~van~der~Horst\altaffilmark{12},
Chryssa~Kouveliotou\altaffilmark{7},
Emily~Layden\altaffilmark{2},
Sheila~McBreen\altaffilmark{6},
Sin\'ead~McGlynn\altaffilmark{6},
Veronique Pelassa\altaffilmark{2},
Robert~D.~Preece\altaffilmark{2},
Arne~Rau\altaffilmark{1},
Dave~Tierney\altaffilmark{6},
Colleen~A.~Wilson-Hodge\altaffilmark{12},
Shaolin~Xiong\altaffilmark{2},
George~Younes\altaffilmark{3}
and Hoi-Fung~Yu\altaffilmark{1,13}
}
\altaffiltext{1}{Max-Planck-Institut f\"{u}r extraterrestrische Physik, Giessenbachstrasse 1, 85748 Garching, Germany}
\altaffiltext{2}{University of Alabama in Huntsville, 320 Sparkman Drive, Huntsville, AL 35805, USA}
\altaffiltext{3}{Universities Space Research Association, 320 Sparkman Drive, Huntsville, AL 35805, USA}
\altaffiltext{4}{Istituto Nazionale di Fisica Nucleare, Sezione di Trieste, I-34127 Trieste, Italy}
\altaffiltext{5}{Dipartimento di Fisica, Universita' di Trieste, I-34127 Trieste, Italy}
\altaffiltext{6}{School of Physics, University College Dublin, Belfield, Stillorgan Road, Dublin 4, Ireland}
\altaffiltext{7}{Astrophysics Office, ZP 12, NASA-Marshall Space Flight Center, Huntsville, AL 35812, USA}
\altaffiltext{8}{NASA Postdoctoral Program, USA}
\altaffiltext{9}{Jacobs Technology, Inc., Huntsville, Alabama}
\altaffiltext{10}{NASA-Goddard Space Flight Center, Greenbelt, MD 20771, USA}
\altaffiltext{11}{Planetarium S\"{u}dtirol, Gummer 5, 39053 Karneid, Italy}
\altaffiltext{12}{Astronomical Institute, University of Amsterdam, Science Park 904, 1098 XH Amsterdam, The Netherlands}
\altaffiltext{13}{Excellence Cluster Universe, Technische Universit\"at M\"unchen, Boltzmannstr. 2, 85748, Garching, Germany}

\begin{abstract}

This is the second of a series of catalogs of gamma-ray bursts (GRBs) observed with the \Fermi\ Gamma-ray Burst Monitor (GBM).  It extends the first two-year catalog by two more years, resulting in an overall list of 953 GBM triggered GRBs.  The intention of the GBM GRB catalog is to provide information to the community on the most important observables of the GBM detected GRBs.  For each GRB the location and main characteristics of the prompt emission, the duration, peak flux and fluence are derived. The latter two quantities are calculated for the 50  -- 300~keV energy band, where the maximum energy release of GRBs in the instrument reference system is observed and also for a broader energy band from 10 -- 1000~keV, exploiting the full  energy range of GBMs low-energy detectors. Furthermore, information is given on the settings and modifications of the triggering criteria and exceptional operational conditions during years three and four in the mission. This second catalog is an official product of the \Fermi\ GBM science team, and the data files containing the complete results are available from the High-Energy Astrophysics Science Archive Research Center (HEASARC).

\end{abstract}


\section{Introduction}

The \Fermi\ Gamma-ray Burst Monitor (GBM), the secondary instrument onboard the \Fermi\ Gamma-ray space Telescope (FGST), launched on 2008 June 11, is now operating successfully in space since five years. GBM's main task is to augment the mission's capability to detect and coarsely locate gamma-ray bursts (GRBs) and to provide broad spectral information.  The GBM instrument extends the energy range of the main instrument, the Large Area Telescope (LAT: 30~MeV -- 300~GeV) down to the soft gamma-ray and X-ray energy range (8~keV -- 40~MeV). This allows for observations over more than seven decades in energy.

In the first four years of operation since triggering was enabled on 2008 July 12, GBM has triggered 2126 times on a variety of  transient events:  954 of these are classified as GRBs (in one case the same GRB triggered GBM twice), 187 as bursts from soft gamma repeaters (SGRs), 261 as terrestrial gamma-ray flashes (TGFs), 394 as solar flares (SFs), 207 as charged particle (CPs) events, and 123 as other events (Galactic sources, accidental statistical fluctuations, or too weak to classify). Table \ref{trigstat1st2ndcat} is a breakdown of the observed event numbers sorted by the time periods covered by the first GBM burst catalog: 2008 July 12 to 2010 July 11 and the additional two years included in the current second catalog: 2010 July 12 to 2012 July 11, separated according to the event type. In addition the numbers of Autonomous Repoint Requests (ARRs, described in Section \ref{SecTriggDiss} below) and GRBs detected by LAT, observed with high confidence above 100 MeV (and 20 MeV), are given \citep{2013ApJS..209...11A}. This catalog lists for each GRB the location and the main characteristics of the prompt emission, the duration, peak flux and fluence. Moreover the distributions of these derived quantities are presented.

The accompanying second spectral catalog \citep[][submitted]{Gru13} provides information on the systematic spectral analysis of nearly all GRBs listed in the current catalog. Time-integrated fluence and peak flux spectra are presented for all GRBs. A catalog reporting time resolved spectral analysis of bright GRBs will be published later (Yu et al., in preparation).
Detailed studies of various GBM GRB subsamples have been presented elsewhere \citep[e.g.][]{Ghirl10,2010ApJ...725..225G, Lv10, 2011ApJ...733...97B, Ghirl11,2011A&A...531A..20G, Nava11a, Nava11b, Zhang11,2012ApJ...756..112L,2013ApJ...763...15Q,2013PASJ...65L...3T}.

In section 2 we briefly introduce the GBM detectors and the GBM GRB localization principle together with a description of the onboard triggering system and path of trigger information dissemination. Furthermore the GBM data products are presented. Section 3 reports the GRB trigger statistics of the first four years, comparing them with the triggers on other event classes. Exceptional operation conditions occurring during years 3 $\&$ 4 are also mentioned. A summary on the major steps of the catalog analysis is given in section 4. The catalog results are presented in section 5 and are discussed in section 6. Finally, in Section 7 we conclude with a summary.

\section{The Gamma-ray Burst Monitor}

\subsection{Burst Detectors}

The ability of the GBM to observe GRBs in the energy range of the maximum energy release in the instrument reference system and to provide energy coverage up to energies of the main instrument is achieved by employing two different kinds of scintillation detectors.
In the energy range from 8~keV to 1~MeV, sodium iodide detectors (NaI) read out by a photomultiplier tube (PMT) are adopted. The capability to coarsely determine locations of triggered GRBs over the full unocculted sky is obtained by using twelve disk shaped NaI crystals, 12.7 cm  diameter by 1.27 cm thick, each of which have a quasi-cosine response, and by arranging the NaI detectors around the spacecraft in such a way that each detector is observing the sky at a different inclination. The location of a GRB is calculated by comparing the measured individual detector counting rates with a lookup table, containing a list of relative detector rates for a grid of simulated sky locations.
The on-board and on-ground lookup tables have resolutions of 5 degrees and 1 degree, respectively. With this method the limiting accuracy is approximately 8 degrees for on-board locations and approximately 4 degrees for on-ground locations. A detailed investigation of the GBM location accuracy can be found in \cite[][submitted]{Con13}.
For the detection of the prompt gamma-ray emission in the MeV-range, between $\sim 200$~keV and 40 MeV, detectors employing the high Z high density scintillation material Bismuth Germanate (BGO) are used. Two detectors using large cylindrical BGO crystals, 12.7 cm diameter by 12.7 cm thick, each viewed by two photomultipliers, are mounted on opposite sides of the spacecraft, allowing observations of the full unocculted sky and providing spectral information up to the MeV regime for all GBM detected bright and hard GRBs. The GBM instrument is described in more detail in \citet{2009ApJ...702..791M}.

\subsection{Trigger Dissemination and Data Products}
\label{SecTriggDiss}

The GBM trigger algorithms implemented in the flight software (FSW) monitor the background count rates of all NaI detectors for the occurrence of a significant count rate increase in different energy ranges and timescales with adjustable sensitivities. A trigger is only generated in case of simultaneous exceedance of the trigger threshold of  at least two detectors, thus reducing the probability for false triggers. The concept of the trigger algorithm was adopted from the predecessor instrument, the Burst and Transient Source Experiment (BATSE) on the {\it Compton Gamma-ray Observatory} ({\it CGRO}), but with advancement in the number of parallel running algorithms. Compared to the BATSE FSW which allowed only for 3 algorithms, running at different timescales in one commandable energy band \citep{1999ApJS..122..465P},
the GBM FSW supports up to 119 trigger algorithms, 28 of which are currently in use. The parameters of all algorithms, i.e., integration time, energy channel range and time offset (see below), are adjustable by command.
With the large number of algorithms and flexibility it is possible to investigate if the population of BATSE observed GRBs was eventually biased by the limited number of trigger algorithms. In addition, the capability was added to run a copy of a search algorithm which is offset in time compared to the original algorithm. From this an improvement in the trigger sensitivity \citep{2002ApJ...578..806B,2004AIPC..727..688B} is expected. The standard setting of the offset is half the timescale of the original algorithm.  A summary of the actual settings (by July 2012) and the changes in the first four years of the mission is shown in Table \ref{trigger:criteria:history}.

Since GBM triggers on events with a broad range of origins in addition to GRBs, the FSW performs an automatic event classification by using a Bayesian approach that considers the event localization, spectral hardness, and the spacecraft geomagnetic latitude \citep{2009ApJ...702..791M}.  This information is very important and useful for the automated follow up observations. Furthermore the capabilities of the instrument to detect events other than GRB events were improved by tuning dedicated trigger algorithms.

In case of a trigger the most important parameters for rapid ground based observations, i.e., onboard localization, event classification, burst intensity and background rates, are downlinked as TRIGDAT data by opening a real-time communication channel through the Tracking and Data Relay Satellite System (TDRSS). In addition, these data are used in near real-time by the Burst Alert Processor (BAP), redundant copies of which are running at the Fermi Mission Operations Center (MOC) at GSFC and the GBM Instrument Operations Center (GIOC) at the National Space Science and Technology Center (NSSTC) in Huntsville, Alabama. Relative to the GBM FSW, the BAP provides improved locations, since it uses a finer angular grid (1 degree resolution) and accounts for differences in the burst spectra and more accurately for atmospheric and spacecraft scattering. Users worldwide are quickly informed within seconds about the flight and automatic ground locations and other important parameters by the automatic dissemination of notices (see Table \ref{GCNnoticetypes} for the different kind of GBM notices) via the GRB Coordinates Network (GCN, \url{http://gcn.gsfc.nasa.gov/}).
The GBM burst advocates (BA), working in alternating 12 hr shifts at the GIOC and at the operations center MGIOC at the Max Planck Institute for Extraterrestrial Physics (MPE) in Garching, Germany, use the TRIGDAT data to promptly confirm the event classification and generate refined localizations by applying improved background models.  Unless a more precise localization of the same GRB has been reported by another instrument, a GCN notice with the final position and classification is disseminated by the BA. In addition the BAs compute preliminary durations, peak fluxes, fluences and spectral parameters, and report the results in a GCN circular in case of a bright event or a GRB that was already detected by another instrument.
Trigger times and locations of GBM triggered GRBs are also passed directly to the LAT in order to launch dedicated onboard burst search algorithms for the detection of accompanying
high-energy emission.
In case of a sufficiently intense GRB, which exceeds a specific threshold for peak flux or
fluence, a request for an autonomous repoint (ARR) of the spacecraft is transmitted to the
LAT and forwarded to the spacecraft.
This observation mode maintains the burst location in the LAT field of view for an extended duration (currently 2.5 hours, subject to Earth limb constraints), to search for delayed high-energy emission. Table \ref{trigstat1st2ndcat}  lists the number of ARRs which occurred in the first four years.

The continuous background count rates, recorded by each detector are downlinked as two complementary data types, the 256 ms high temporal resolution CTIME data with 8 energy channels and the 4 s low temporal resolution CSPEC data with full spectral resolution of 128 energy channels which is used for spectroscopy.
The lookup tables (LUTs) used to define the boundaries of the CTIME and CSPEC  spectral energy channels are pseudo-logarithmic so that the widths are commensurate with the detector resolution as a function of energy. In case of an on-board trigger the temporal resolutions of CTIME and CSPEC data are increased to 64~ms and 1~s, respectively, a mode lasting nominally for 600 s after the trigger time.

Moreover, high temporal and spectral resolution data are downlinked for each triggered event. These time-tagged event (TTE) data consist of individually recorded pulse height events with 2~$\mu$s temporal resolution and 128 channel spectral resolution from each of the 14 GBM detectors, recorded for 300~s after and about 30~s before trigger time. The benefit of this data type is the flexibility to adjust the temporal resolution to an optimal value with sufficient statistics for the analysis in question.

\section{In-Orbit Operations}

\subsection{Trigger Statistics}
\label{trigger_statistics}

\begin{figure}
\begin{center}
\includegraphics[scale=0.75]{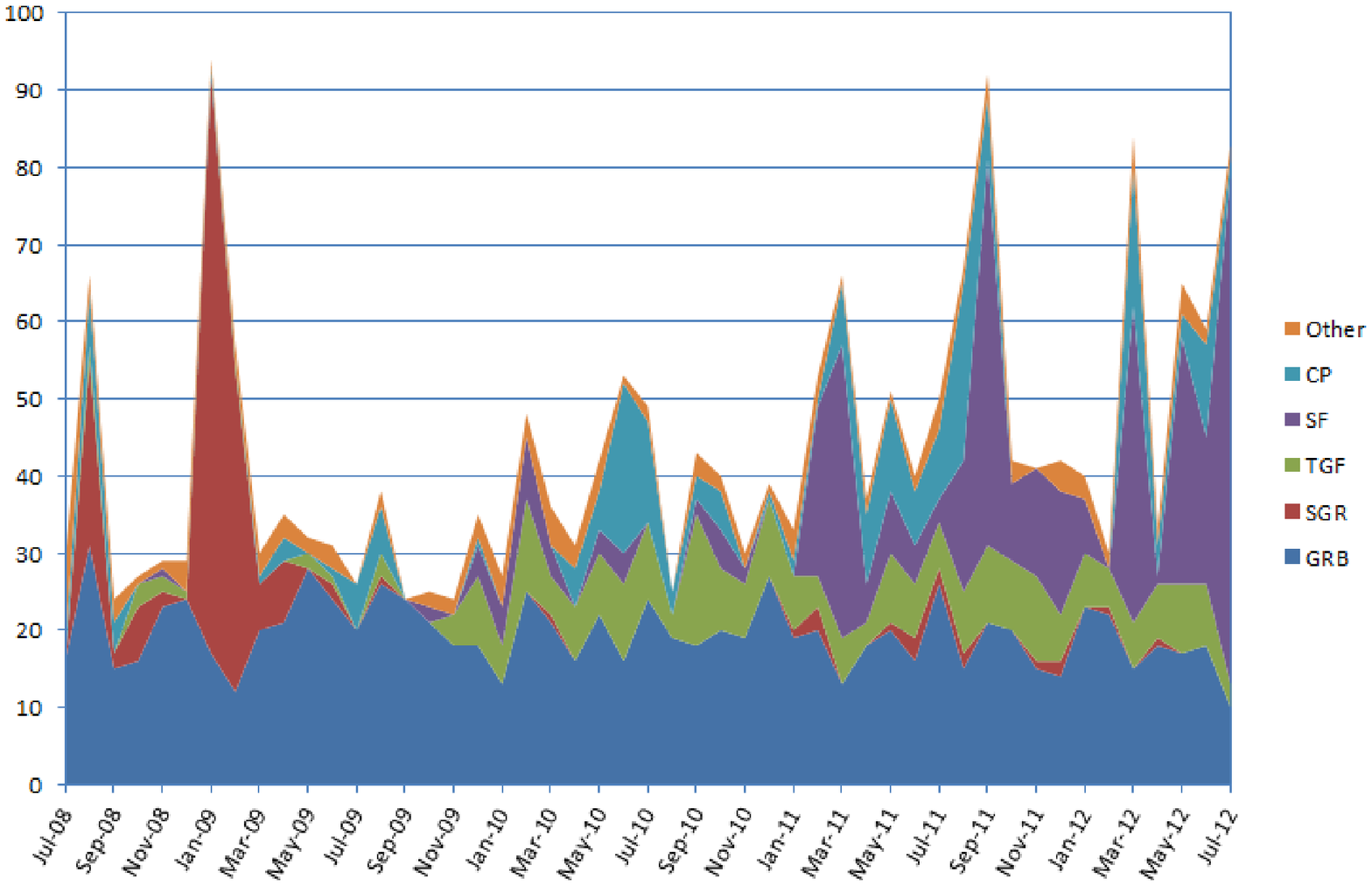}
\caption{\label{monthlytrigstat} The monthly trigger statistics over the first four years of the mission. For 2008 July and 2012 July only the number of triggers in the time period from 2008 July 12 to 31 and 2012 July 1 to 11 are shown.}
\end{center}
\end{figure}

The GBM instrument, which is primarily designed to detect cosmic GRBs, additionally detects bursts originating from other cosmic sources, such as SFs and SGRs, as well as extremely short but spectrally hard TGFs observed from the Earth's atmosphere, which have been associated with lightning events in thunderstorms. Table \ref{trigstat1st2ndcat}  summarizes the numbers of triggers assigned to these additional event classes, showing that their total number is of the same order as the total number of triggered GRBs. Approximately 10\% of the triggers are due mostly to cosmic rays or trapped particles; the latter typically occur in the entry region of the South Atlantic Anomaly (SAA) or at high geomagnetic latitude. In rare cases outbursts from known Galactic sources have caused triggers. Finally, $\sim 6$\% of the GBM triggers are generated accidentally by statistical fluctuations or are too weak to be confidently classified.
The monthly trigger statistics over the first four years of the mission is graphically represented in Figure \ref{monthlytrigstat}. The rate of GRBs is slightly lower in the second two years because at the beginning in 2011 July triggers were disabled during times when the spacecraft was at high geomagnetic latitude. It is evident from Figure~\ref{monthlytrigstat} that the major bursting activity from SGR sources took place in the beginning of the mission, mainly in 2008 and 2009. In addition to emission from previously known SGR sources \citep{2012ApJ...755..150V,2012ApJ...749..122V,2011ApJ...739...87L}, GBM also detected a new SGR source \citep{2010ApJ...711L...1V}.
It is also obvious from the figure that the rate of monthly detected triggers on TGF events has increased by a factor of $\sim 8$ to about two per week, after the upload of the new  FSW version  on 2009 November 10 \citep{2011JGRA..116.7304F}. This version includes additional trigger algorithms that monitor the detector count rates of the BGO detectors in the 2 -- 40~Mev energy range (see Table \ref{trigger:criteria:history}). This is advantageous because the TGF bursts show very hard spectra up 40~MeV,  which also increases the deadtime in the NaI detectors \citep{2013JGRA..118.3805B}.

Table \ref{trigstatalgor} summarizes which trigger algorithm has triggered first on bursts or flares from the different object classes. Once a trigger has occurred the FSW continues to check the other trigger algorithms and ultimately sends back the information in TRIGDAT data as list of trigger times for all algorithms that triggered. This detailed information was already used in \cite{Pacie12} to investigate the apparent improvement in trigger sensitivity relative to BATSE.
A breakdown of GBM GRBs which triggered on BATSE- and non-BATSE-like trigger algorithms, individually listed for the first and second catalog periods is shown in Table \ref{grbbreakdown}.
It was found that mainly GBM's additional longer trigger timescales triggers ($> 1.024$~s) in the 50 to 300 keV energy range were able to detect GRB events which wouldn't have triggered the BATSE experiment. These observations are confirmed by analyzing the current full 4 year dataset.
Furthermore we ascribe the improved trigger sensitivity to the in general lower trigger threshold of $4.5\sigma - 5.0\sigma$ (see Table \ref{trigger:criteria:history}) compared to the BATSE settings  \citep[see Table 1 in][]{1999ApJS..122..465P}.
The longest timescale trigger algorithms in the 50 - 300~keV energy range, running at $\sim 16$~s (20, 21) and $\sim 8$~s (18, 19) were disabled in the beginning of the mission (see Table \ref{trigger:criteria:history}), since no  event triggered algorithms 20 \& 21 and only three GRBs algorithms 18 \& 19.
The algorithms running on energy channel 2 (25 -50 keV) with timescales higher than 128~ms were disabled, since they were mostly triggered by non GRB (and non SGR) events.
The short timescale algorithms in the 25 - 50~keV energy range (22 - 26) were kept,
mainly for the detection of SGR bursts, which are short and have soft energy spectra.
The new algorithms above 100~keV  didn't increase the the GRB detection rate. They were disabled with the exception of the shortest timescale algorithms running at 16~ms, particulary suitable for the detection of  TGFs. Table \ref{trigger:criteria:history} clearly shows the capabilities of the newly introduced "BGO"-trigger algorithm 116 -119 for TGF detection.

\subsection{Exceptional Operational Conditions:  Year Three and Four}

At various times during years 3 \& 4 the instrument configuration was temporarily changed in two ways that affect the GRB data: 1) some or all of the trigger algorithms were disabled, and 2) the low-level energy thresholds (LLT) were raised on the sun-facing detectors (NaI 0-5)\footnote{A table summarizing the intervals of non-nominal LLT settings is posted at:\\ \url{http://fermi.gsfc.nasa.gov/ssc/data/access/gbm/llt\_settings.html}}.
It is evident in Figure \ref{monthlytrigstat} that the number of triggers due to SFLs increased significantly around the beginning of 2011, an effect which is consistent with entering an active phase in the 11-year solar cycle. Solar flares typically have very soft spectra, producing high rates of low energy events in the solar-facing GBM detectors. Due to concerns that this might result in an unacceptable amount of TTE data, the LLTs were raised during two intervals when solar activity was high, a solar Target of Opportunity pointing on 2011 Sept 8 - 10 and an interval of high solar activity beginning on 2012 July 11 and continuing beyond the end of the period covered by this catalog.

The potential for high rates of soft solar X-rays was also a concern for a series of nadir pointings intended to detect TGFs in the LAT. During those intervals all triggering was disabled but TTE generation was turned on continuously so that a sensitive search for GBM TGFs coincident with the LAT could be performed. Again, in order to mitigate against unacceptably high rates of TTE, the LLTs in the sun-facing NaI detectors were raised above the nominal.

GBM TTE data suffer from timing glitches arising from rare conditions in the FPGA logic that produces GBM science data onboard. Every effort is made on the ground to correct these glitches but some are not cleanly reparable using pipeline software logic, and the TTE data files occasionally show the effects of these glitches, which can be seen in the TTE lightcurves\footnote{More details are provided at \url{http://fermi.gsfc.nasa.gov/ssc/data/analysis/GBM\_caveats.html}}. During pointed Target of Opportunity observations of the Crab Nebula between 2012 July 5 - 9, the on-board electronics were subjected to unusually low temperatures that caused a higher than normal rate of the TTE timing glitches.

The success of GBM in detecting TGFs led to great efforts to further increase the number of detected events \citep{2013JGRA..118.3805B}. A fundamental hardware limitation of the onboard triggering process is the minimum integration time of 16~ms. This integration time is much longer than the duration of a typical TGF of about 0.1~ms, which adds unnecessary background data and reduces the trigger sensitivity. These limitations are circumvented by downlinking the GBM photon data as continuous TTE data and by conducting a ground-based search for TGFs. In order to limit the data volume the GBM photon TTE data were only gathered over select parts of orbit (moving boxes) where the highest seasonal thunderstorm activity is expected. This mode was first implemented on 2010 July 15 and is typically enabled for $\sim 20$\% of the observing time. The resulting increase of the TGF detection rate is a factor of 10 compared to the rate of TGF triggers.

\section{GRB Catalog Analysis}

The GBM GRB catalog analysis process is described in detail in the first catalog paper (see appendix of \cite{Pacie12}). Here we present and summarize the major analysis steps for the better understanding of the presented results and GRB tables.
The analysis results for each burst of the current catalog were discussed in detail within the GBM catalog team and confirmed in case of consensus. In several cases a reanalysis was necessary.  This validation procedure ensures the compliance of the results with the GRB selection and analysis criteria which are defined by the GBM science team.

\subsection{Burst Localization and Instrument Response}

The GRB locations listed in Table \ref{main_table} are adopted from the BA analysis results, uploaded to the GBM trigger catalog at the NSSTC (with a copy at the FSSC). Non-GBM locations are listed for bursts that were detected by an instrument providing a better location accuracy, such as \Swift\ BAT \citep{2005SSRv..120..143B} or XRT \citep{2005SSRv..120..165B}, or were localized more precisely by the Inter Planetary Network \citep[IPN,][]{2013ApJS..207...39H}.

For each GRB the individual detector response matrices (DRMs) needed for analysis of the science data were generated for the best location using version GBMRSP v1.9 or v2.0 of the response generator and version 2 of the GBM DRM database. The detector response is dependent on incident photon energy, the measured detector output energy, and the detector-source angle. Two sets of DRMs are generated, one for 8-channel (CTIME) data and one for 128-channel (CSPEC \& TTE) data. The Earth-source-spacecraft geometry is also considered in order to account for contributions from earth's atmosphere scattering.  In case of relatively long duration GRBs RSP2 response files with multiple DRMs are used, which provide a new DRM  every 2$^{\circ}$ of satellite slew.

The determination of a reliable location is quite important since all analysis results  depend on the response files generated for the particular GRB location. Systematic errors of the localizations are evaluated by comparing GBM locations with "true" locations from higher spatial resolution instruments or the IPN \citep[see][submitted]{Con13}.

\subsection{Duration, Peak Flux and Fluence Analysis}

The analysis performed to derive the duration, peak flux and fluence of each burst is based on an automatic batch fit routine implemented within the RMFIT software\footnote{We used the spectral analysis package RMFIT, which was originally developed for time-resolved analysis of BATSE GRB data but has been adapted for GBM and other instruments with suitable FITS data formats. The software is available at the Fermi Science Support Center: \url{http://fermi.gsfc.nasa.gov/ssc/data/analysis/user/}.}.

The data typically used for this analysis are either the CTIME or CTTE\footnote{For the duration analysis of short bursts a dedicated data type can be produced from the TTE data by using a separate software.  It is redistributing the counts of the 128 spectral channels to eight channels, providing a high time resolution CTIME like data type called CTTE. Since the TTE data are available for about 30~s pretrigger the derived CTTE data type is also suitable for the analysis of bursts which show their peak-flux interval during the pretrigger time.} data from those NaI detectors that view the burst with an angle of incidence less than 60 degrees, without significant blockage by the spacecraft or LAT components. Detectors may be omitted if they have rapidly varying backgrounds (e.g., due to solar activity).
If available, CTTE data are used if the burst is too short to resolve with CTIME or if the peak is before the trigger time. The energy range is set to $\sim 10$~keV to $\sim 1$~MeV by selecting all but the first and last energy channels. The temporal resolution is typically set to 256~ms, but may be as short as 64~ms if necessary to resolve very short bursts.  Source and background time intervals are then selected.  The source interval covers the burst emission time plus approximately equal intervals of background before and after the burst (generally at least $\sim 20$ seconds on either side). Background intervals are selected before and after the burst, about twice as wide as the burst emission and having a good overlap with the source interval. In the case of a burst with quiescent times between pulses, additional background intervals may be selected within the source interval.  RMFIT computes a background model by fitting a polynomial of up to 4$^{\rm th}$ order to the selected background intervals, separately for each detector and energy channel. Depending on the background variability, the lowest order polynomial that gives a good fit is selected.

RMFIT then deconvolves the counts spectrum of each time bin in the source interval, yielding a photon flux history over the selected energy range. For this analysis, the "Comptonized" (COMP) photon model was used:
\begin{displaymath}
f_{COMP}(E) = A \ \Bigl(\frac{E}{E_{piv}}\Bigr) ^{\alpha} \exp \Biggl[ -\frac{(\alpha+2) \ E}{E_{peak}}  \Biggr],
\end{displaymath}
characterized by the parameters: amplitude \emph{A}, the low energy spectral index $\alpha$ and peak energy $E_{peak}$ (the parameter $E_{piv}$ is fixed to 100 keV).

\begin{figure}
\includegraphics[scale=0.60,angle=90.0]{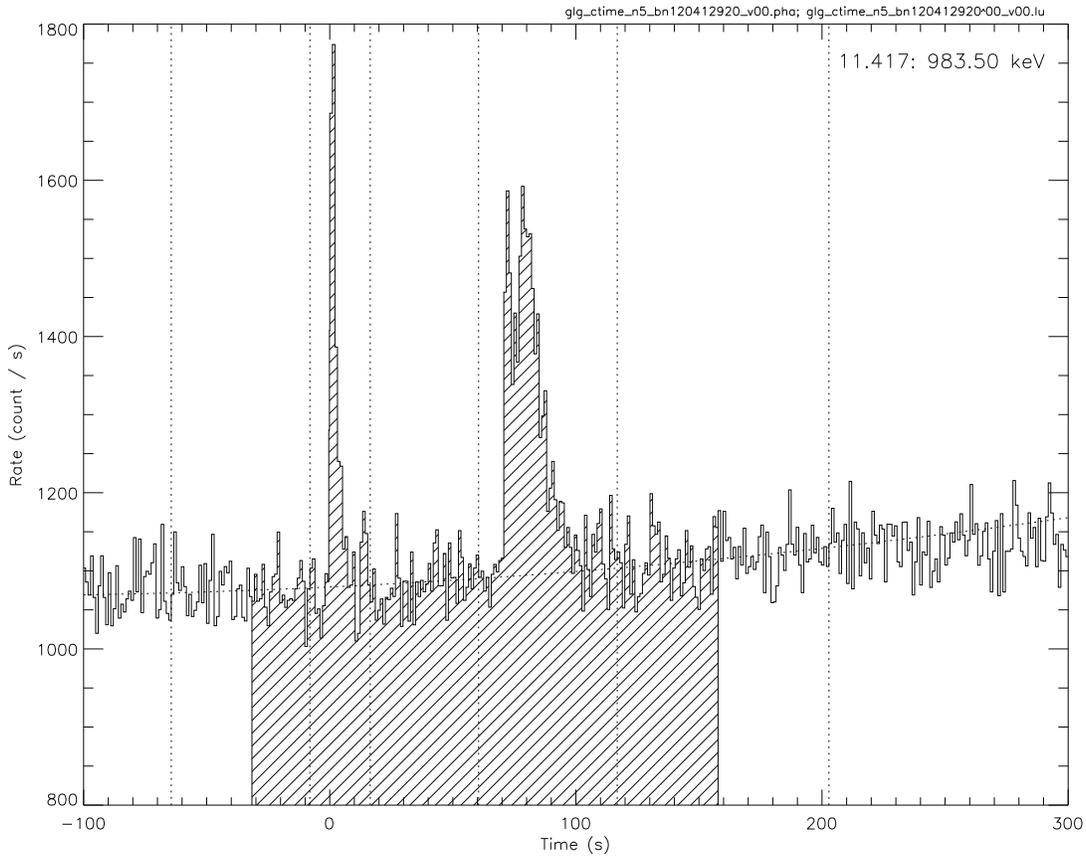}
\caption{\label{examplelc} CTIME lightcurve of GRB 120412A (bn120412920) with a 1.024~s temporal resolution in NaI detector~5. Vertical dotted lines indicate the regions selected for fitting the background (in this case three regions). The hatching defines the source region selected for the duration analysis.}
\end{figure}

\begin{figure}
\begin{center}
\epsscale{0.75}
\plotone{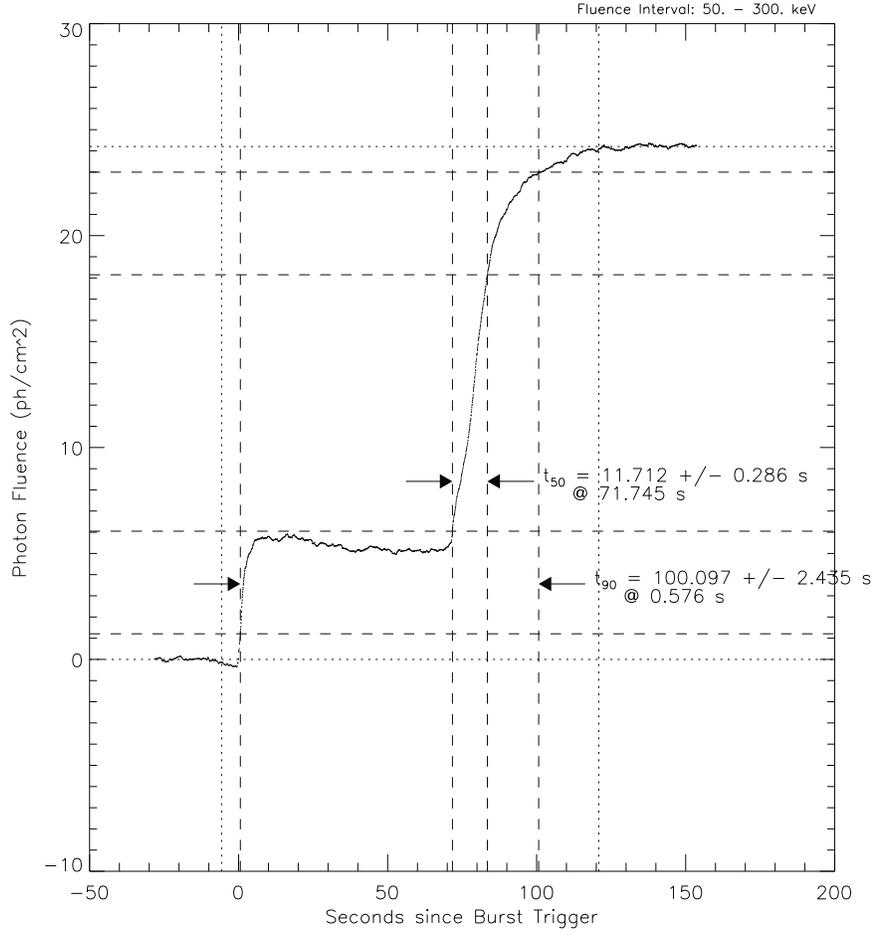}
\caption{\label{exampledur} The duration plot for GRB 120412A (bn120412920) is an example of the analysis for a GRB showing two emission periods separted by a longer quiecent time interval. Data from NaI detectors 2 \& 5 were used. Horizontal dotted lines are drawn at 5\%, 25\%, 75\% and 95\% of the total fluence. Vertical dotted lines are drawn at the times corresponding to those same fluences, thereby defining the $T_{50}$ and $T_{90}$ intervals.}
\end{center}
\end{figure}

Figure \ref{examplelc} shows the light curve as measured by a single NaI detector of a relatively long burst consisting of two major emission periods, with the selected source and background intervals highlighted.  Figure \ref{exampledur} shows a plot of the increase of the integrated flux in the 50 -- 300~keV energy range derived from the model fitting for all time bins within the source interval. The three plateaus are the time intervals where no burst emission is observed. This function is used to determine the T50 (T90) burst duration from the interval between the times where the burst has reached 25\% (5\%) and 75\% (95\%) of its fluence, as illustrated by the horizontal and vertical dashed lines.

Peak fluxes and fluences are obtained in the same analysis, using the same choices of detector subset, source and background intervals, and background model fits. The peak flux is computed for three different time intervals: 64 ms, 256 ms and 1.024 s in the energy range 10 -- 1000~keV and, for comparison purposes with the results presented in the BATSE catalog \citep{1998AIPC..428....3M}, in the 50 -- 300~keV energy range. The burst fluence is also determined in the same two energy ranges. The RMFIT analysis results presented above are stored in a BCAT fits file: glg\_bcat\_all\_bnyymmddttt\_vxx.fit with specified wildcards for the year (yy), month (mm), day (dd), fraction of a day (ttt) and version number (xx).

\section{Catalog Results}
The catalog results can be accessed electronically through the HEASARC browser interface (\url{http://heasarc.gsfc.nasa.gov/W3Browse/fermi/fermigbrst.html}). Standard light curve plots for each burst can be viewed at \url{http://gammaray.nsstc.nasa.gov/gbm/science/grbs/month\_listings/}. Here we provide tables that summarize selected parameters.

Table~\ref{main_table} lists the 954 triggers of the first four years that were classified as GRBs. The GBM Trigger ID is shown along with a conventional GRB name as defined by the GRB-observing community.
For readers interested in the bursts with significant emission in the BGOs, the trigger ID and GRB name are highlighted in italics if emission in the BGO data (above 300 keV) is visible in the standard light curve plots\footnote{These BGO-detected identifications are the result of a visual search rather than a quantitative analysis and thus do not have a well-defined threshold.}.
Note that the entire table is consistent with the small change in the GRB naming convention that became effective on 2010 January 1 \citep{Barth09}: if for a given date no burst has been ``published'' previously, the first burst of the day observed by GBM includes the 'A' designation even if it is the only one for that day.
The third column lists the trigger time in UT. The next four columns in Table~\ref{main_table} list the sky location and
associated error\footnote{For GBM derived locations the statistical 1-sigma error is given. The GBM errors are not
symmetric and the given value is the average of the error ellipse.} along with the instrument that determined the location.
The table lists the GBM-derived location only if no higher-accuracy locations have been reported by another instrument. The choice of a higher-accuracy location is somewhat arbitrary (e.\,g., \Swift-BAT locations are often listed even if a \Swift-XRT location is available); for the GBM analysis, location accuracy better than a few tenths of a degree provides no added benefit. The table also shows which algorithm was triggered along with its timescale and energy range. Note that the listed algorithm is the first one to exceed its threshold but it may not be the only one. Finally, the table lists other instruments that detected the same GRB\footnote{This information was drawn from the IPN master burst list compiled on 2013 July 9, accessible at \url{http://www.ssl.berkeley.edu/ipn3/masterli.txt} \citep[see also][]{2013ApJS..207...39H} and the \INTEGRAL\ IBIS-ISGRI GRB list, accessible at  \url{http://www.isdc.unige.ch/integral/science/grb\#ISGRI}}.

The results of the duration analysis are shown in Tables~\ref{durations}, \ref{pf_fluence} \& \ref{pf_fluence_b}. The values of $T_{50}$ and $T_{90}$ in the 50--300~keV energy range are listed in Table~\ref{durations} along with their respective 1-sigma statistical error estimates and start times relative to the trigger time.  For a few GRBs the duration analysis could not be performed, either due to the weakness of the event or due to technical problems with the input data. Also, it should be noted that the duration estimates are only valid for the portion of the burst that is visible to GBM. If the burst was partially Earth-occultated or had significant emission while GBM detectors were turned off in the SAA region, the ''true'' durations may be underestimated or overestimated, depending on the intensity and variability of the non-visible emission. Finally, for technical reasons it was not possible to do a single analysis of the unusually long GRB 091024A \citep{2011A&A...528A..15G}, so the analysis was done separately for the two triggered episodes. These cases are all noted in the Table. The reader should also be aware that for most GRBs the analysis used data binned no finer than 64~ms, so the duration estimates (but not the errors) are quantized in units of 64~ms. For a few extremely short events (noted in the table) TTE/CTTE data were used with 32~ms or 16~ms binning.

As part of the duration analysis, peak fluxes and fluences were computed in two different energy ranges. Table~\ref{pf_fluence} shows the values in 10-1000~keV and Table~\ref{pf_fluence_b} shows the values in 50--300~keV.
The analysis results for low fluence events are subject to large systematic errors and should be used with caution\footnote{The fluence measurements in the spectroscopy catalog \citep[][submitted]{Gru13} are more reliable for such weak events.}.

\section{Discussion}

\begin{figure}
\begin{center}
\plotone{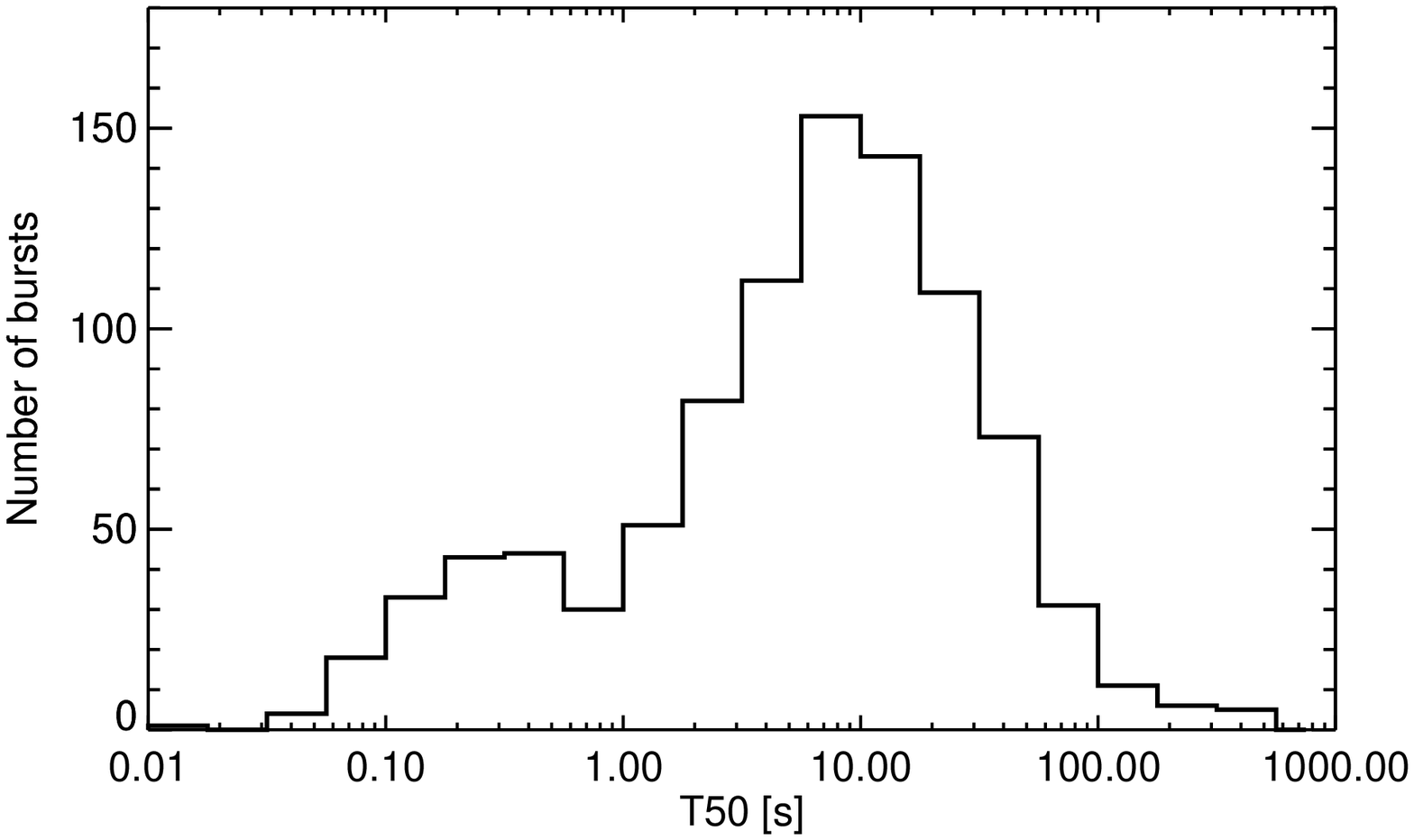}
\plotone{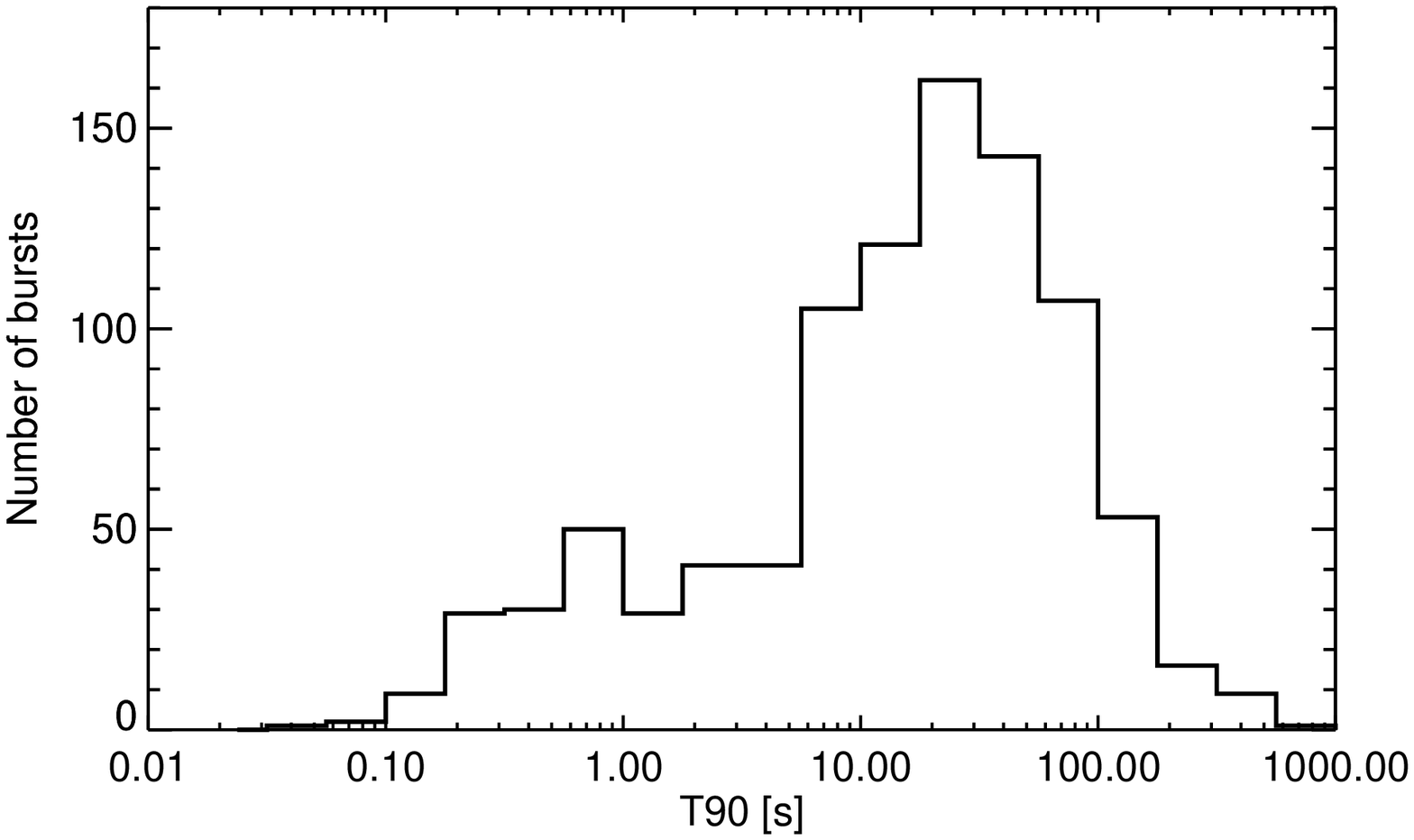}
 \caption{\label{dur_dist} Distribution of GRB durations in the 50--300~keV energy range. The upper plot shows $T_{50}$ and the lower plot shows $T_{90}$. }
\end{center}
 \end{figure}

In the current catalog we are providing the same set of figures as shown in the first catalog. The histograms of the $T_{50}$ and $T_{90}$ distributions are shown in Figure~\ref{dur_dist}. Using the conventional division between the short and long GRB classes of $T_{90} = 2$~s we find for the now longer mission period of four years a slightly lower fraction of short GRBs (see Table \ref{grbbreakdown}). 159 (17\%) of the 953 measured GRBs can be assigned to the short GRB class, within the quoted duration errors, the number ranges from 124 (13\%) to 193 (20\%).
As already claimed in the first catalog we ascribe the lower number of short GRBs observed with GBM compared to BATSE (24\%) not to a deficit of short events but rather to an excess of long events detected by GBM's longer timescale trigger algorithms (see Section \ref{trigger_statistics}).
Furthermore GBM slightly favors triggering on long GRBs,  since the thresholds for the~64 ms timescales are higher ($5.0 \sigma$, see Table \ref{trigger:criteria:history}) than for 256 \& 1024~ms (both $4.5 \sigma$)\footnote{I should be noted that there were also times when BATSE triggers did not use the same threshold for all 3 timescales \citep[see Table 1 in][]{1999ApJS..122..465P}. Selecting periods where all three BATSE trigger algorithm were set to the same value (e.g. a threshold of $5.5 \sigma$ from 1992 September 14 to 1994 September 19 and from 1996 August 29 to the end of the mission) the observed fraction of short GRBs is 24\% (313 short out of 1307  GRBs, see \url{http://heasarc.gsfc.nasa.gov/W3Browse/all/batsegrb.html}).}.
Considering only GRB triggers on one of the two BATSE like algorithms offset half the trigger timescale (see Table \ref{grbbreakdown}) and assuming the best case that all triggers gained by the other algorithm were on long GRBs we derive a slightly higher fraction of short GRBs (e.g. for the full 4 year dataset 19\%). Moreover considering the broad range of the short GRB fraction shown in Table \ref{grbbreakdown}, the GBM BATSE-like trigger fraction on short GRBs comes close to that of the BATSE sample.
We would like to stress that the GBM and BATSE samples of short GRBs are relatively small, so that they are not statistically inconsistent.
From table \ref{grbbreakdown} it emerges that the observed fraction of short GRBs decreased slightly from the first to the second two years, which could be interpreted as a statistical downward fluctuation.

\begin{figure}
\begin{center}
\epsscale{0.9}
\plotone{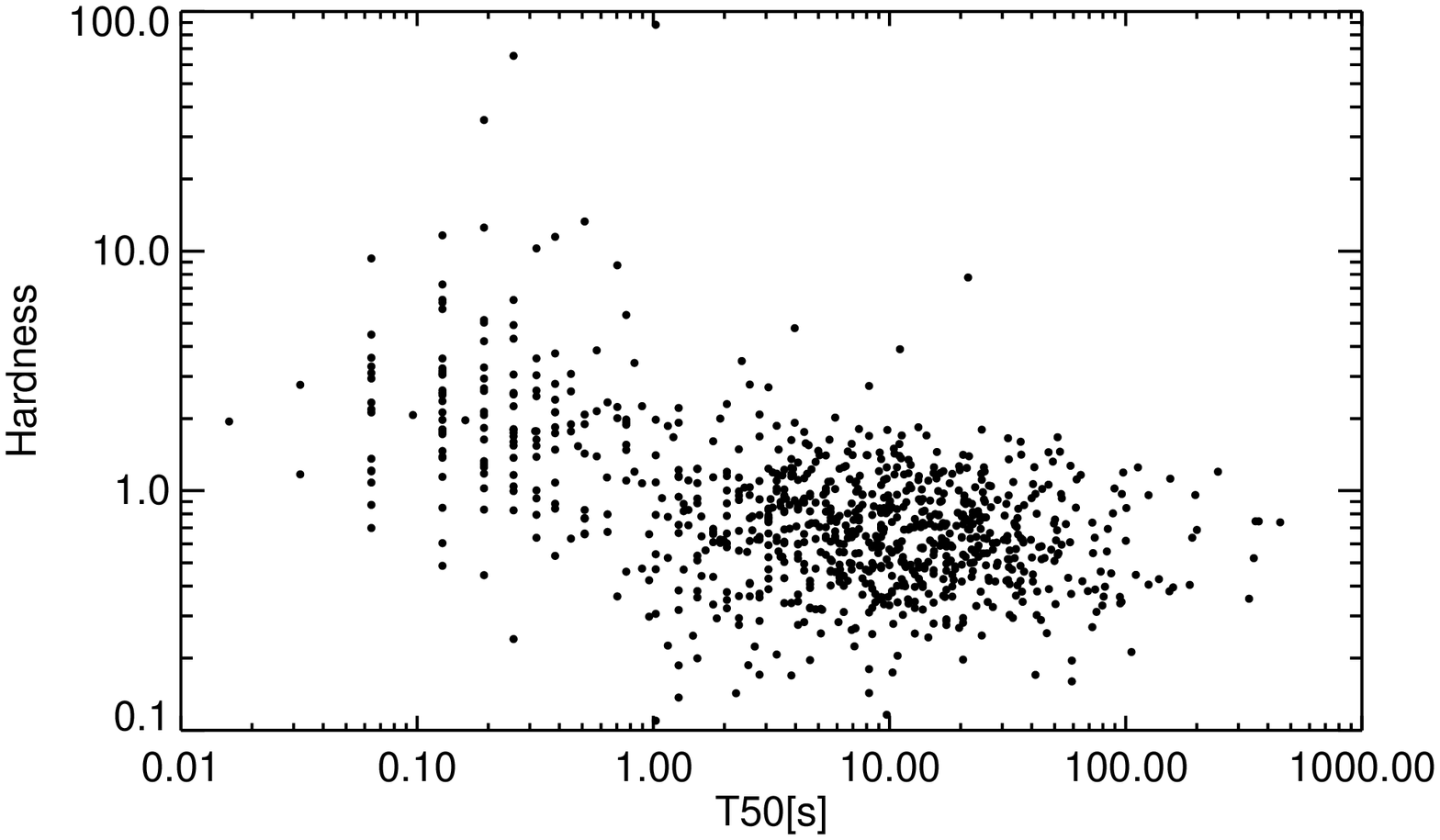}
\plotone{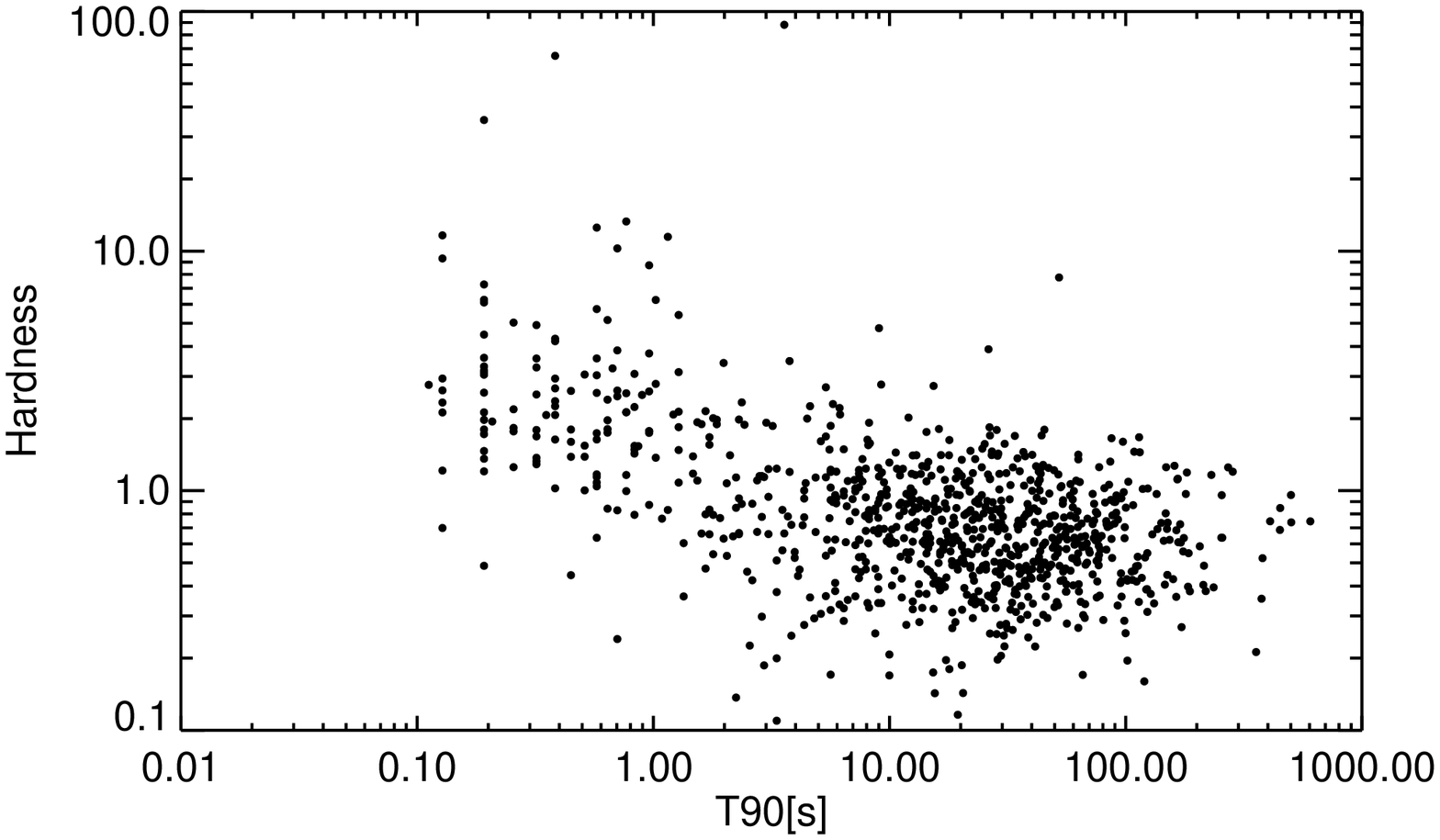}
\caption{\label{hardness_vs_dur} Scatter plots of spectral hardness vs.\ duration are shown for the two duration measures $T_{50}$ (upper plot) and $T_{90}$ (lower plot). The spectral hardness was obtained from the duration analysis results by summing the deconvolved counts in each detector and time bin in two energy bands (10 -- 50~keV and 50 -- 300~keV), and further summing each quantity
in time over the $T_{50}$ and $T_{90}$ intervals.  The
hardness was calculated separately for each detector as the ratio of the flux density
in 50 -- 300 keV to that in 10 -- 50 keV and finally averaged over detectors. For clarity, the estimated errors are not shown but can be quite large for the weak events. Nevertheless, the anti-correlation of spectral hardness with burst duration is evident.}
\end{center}
\end{figure}

Figure~\ref{hardness_vs_dur} shows scatter plots of hardness vs.\  $T_{50}$- and $T_{90}$-durations,  showing that the GBM data are also exhibiting the well known anti-correlation of spectral hardness with duration as known from BATSE data \citep{CK93}. In this analysis the hardness was derived from the time-resolved spectral fits for each GRB, by using the  photon model fit parameters, which are a by-product of the duration analysis.

\begin{figure}
\begin{center}
 \epsscale{0.9}
\plotone{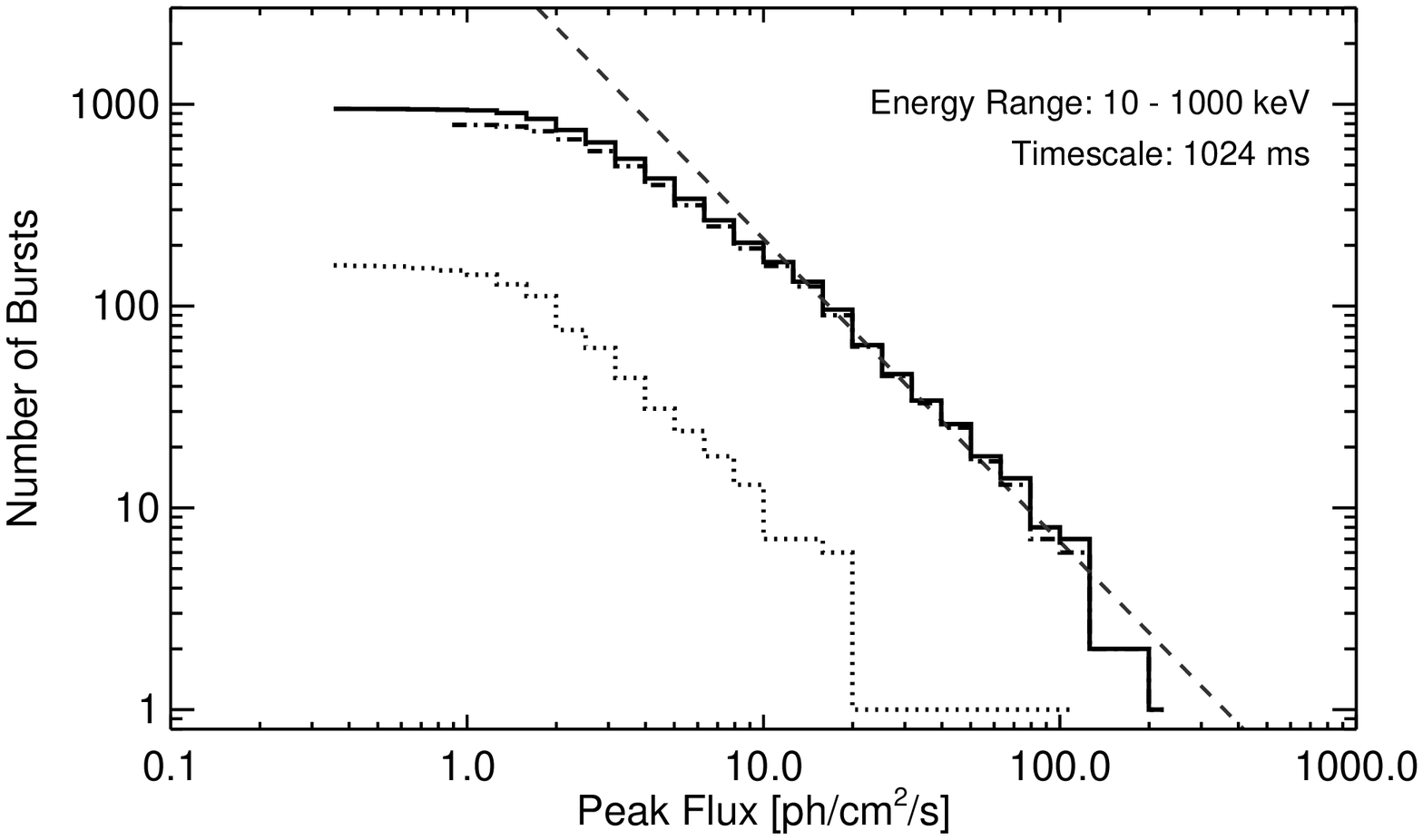}
\plotone{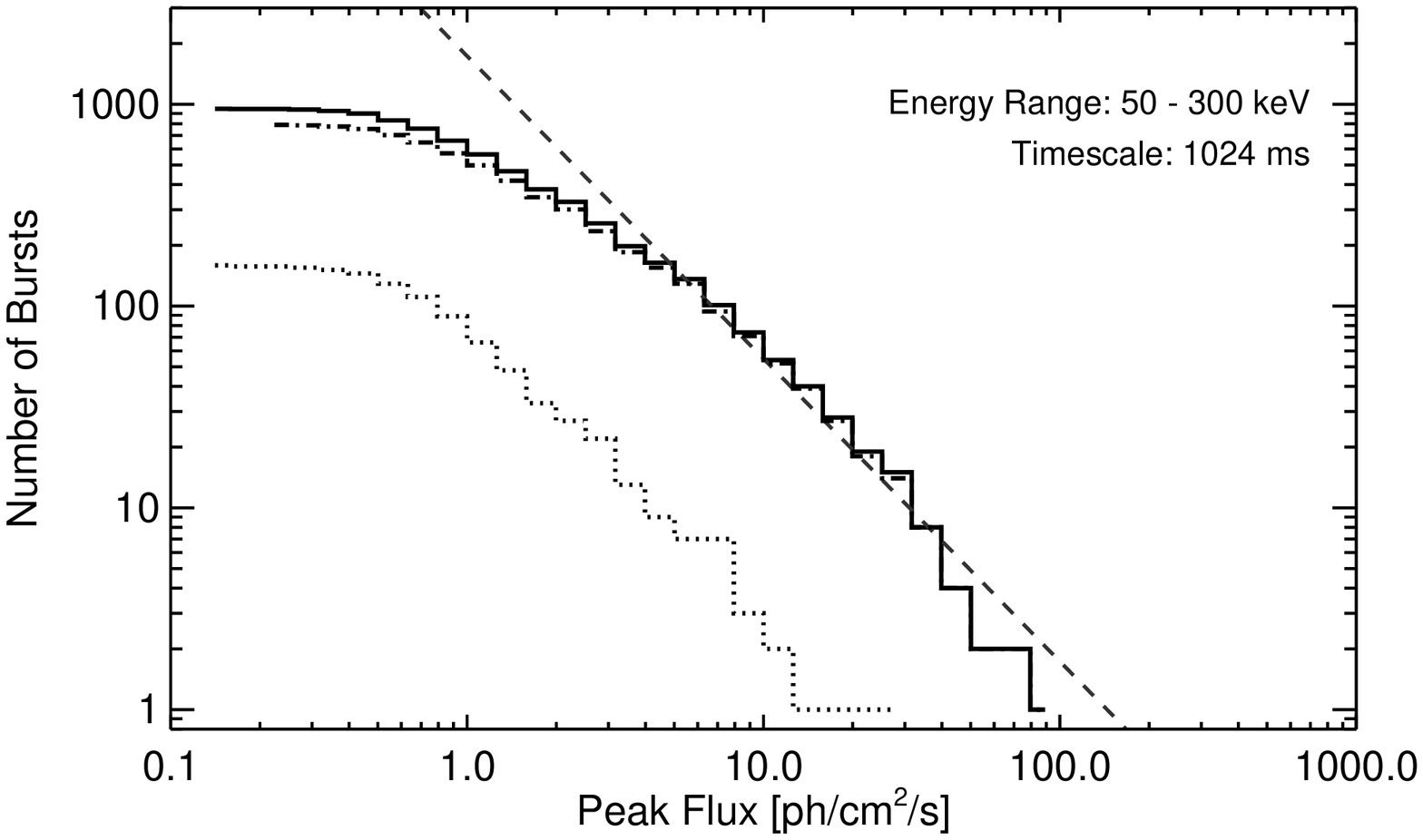}
\caption{\label{pflx_fig} Integral distribution of GRB peak flux on the 1.024~s timescale. Energy ranges are 10 -- 1000~keV (upper plot) and 50 -- 300~keV (lower plot). Distributions are shown for the total sample (solid histogram), short GRBs (dots) and long GRBs (dash-dots), using $T_{90} = 2$~s as the distinguishing criterion. In each plot a power law with a slope of $-3 / 2$ (dashed line) is drawn to guide the eye.}
\end{center}
\end{figure}

\begin{figure}
\begin{center}
 \epsscale{0.9}
\plotone{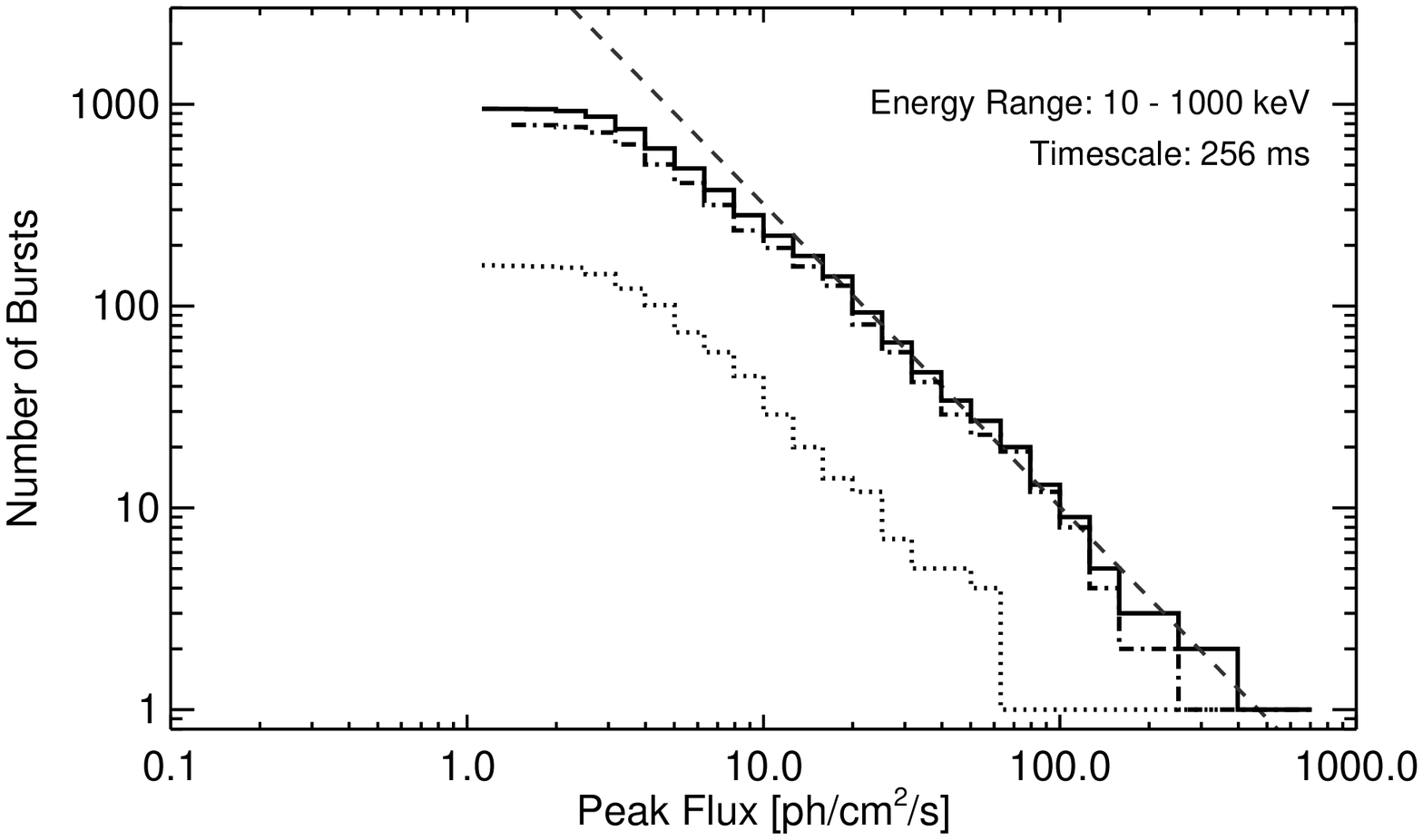}
\plotone{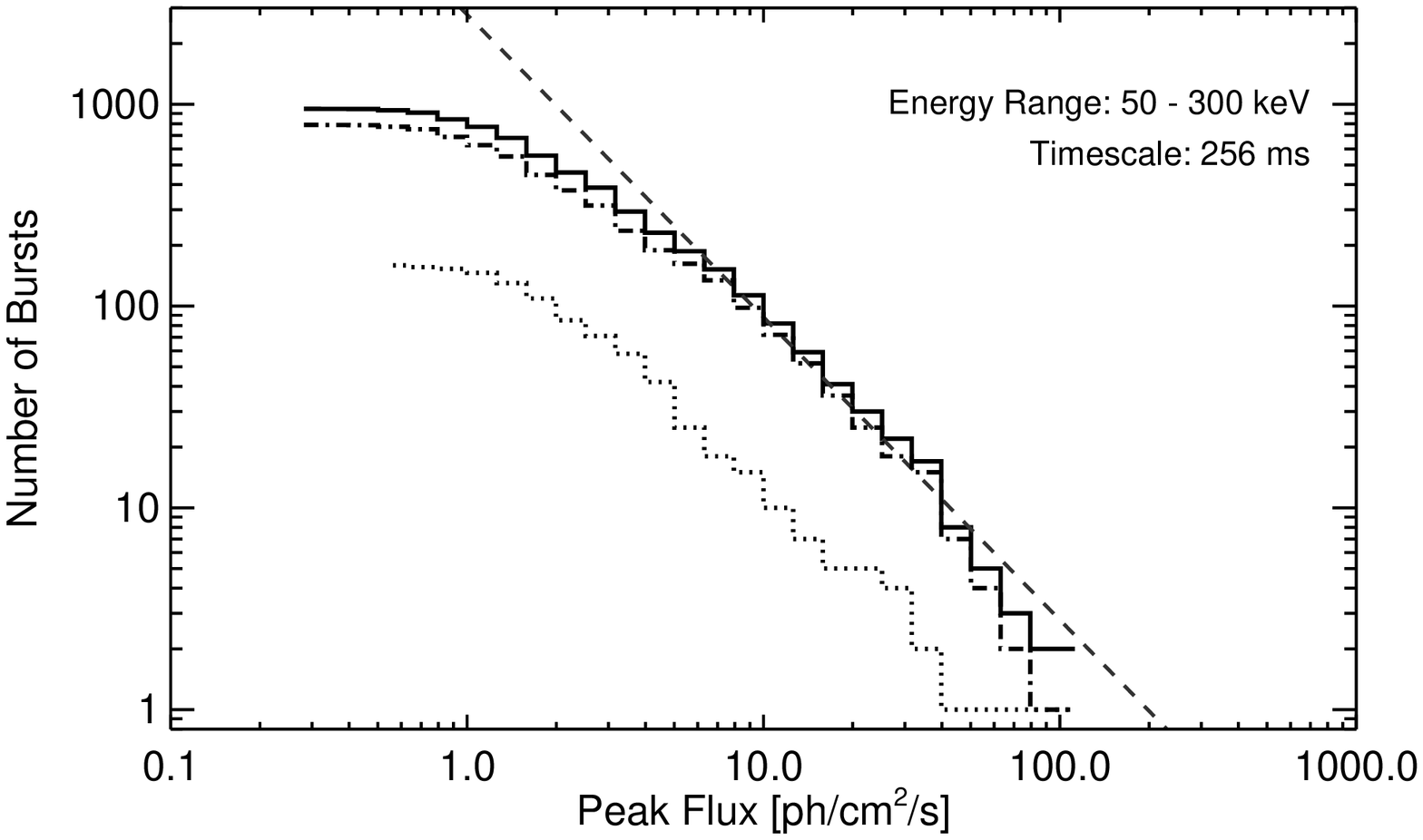}
\caption{\label{pf256_fig} Same as Figure~\ref{pflx_fig}, except on the 0.256~s timescale.}
\end{center}
\end{figure}

\begin{figure}
\begin{center}
 \epsscale{0.9}
\plotone{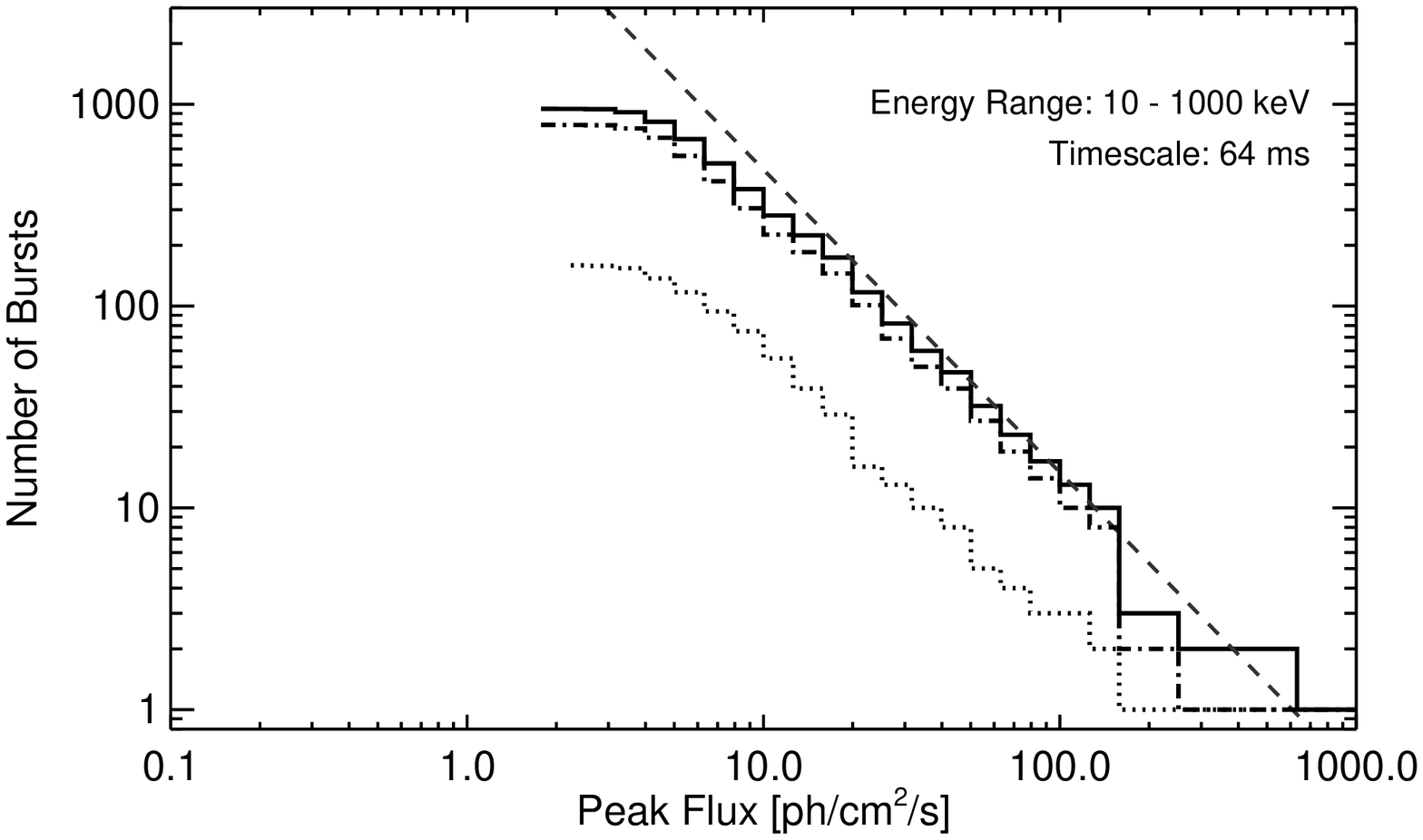}
\plotone{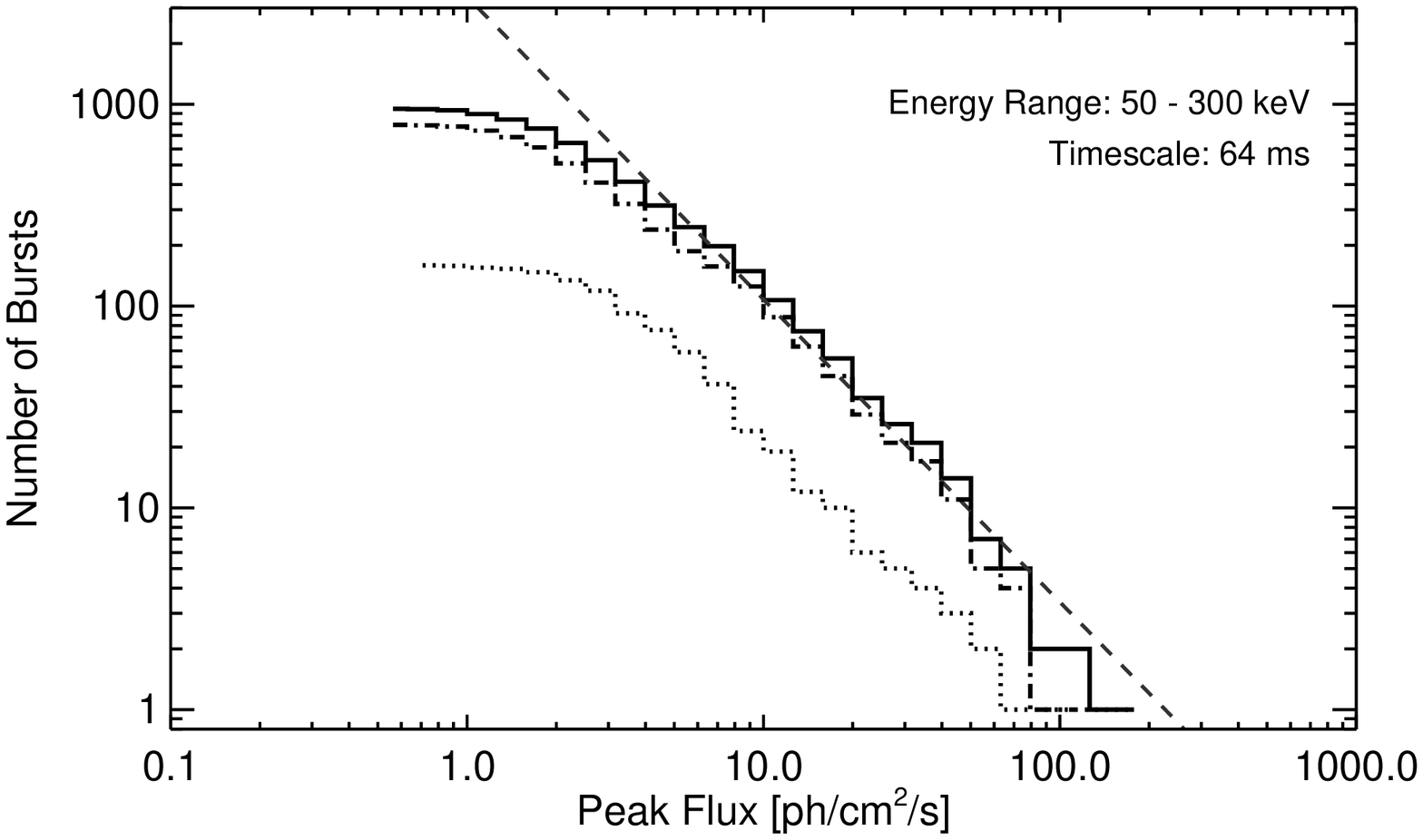}
\caption{\label{pf64_fig}Same as Figure~\ref{pflx_fig}, except on the 0.064~s timescale.}
\end{center}
\end{figure}

\begin{figure}
\begin{center}
 \epsscale{0.8}
\plotone{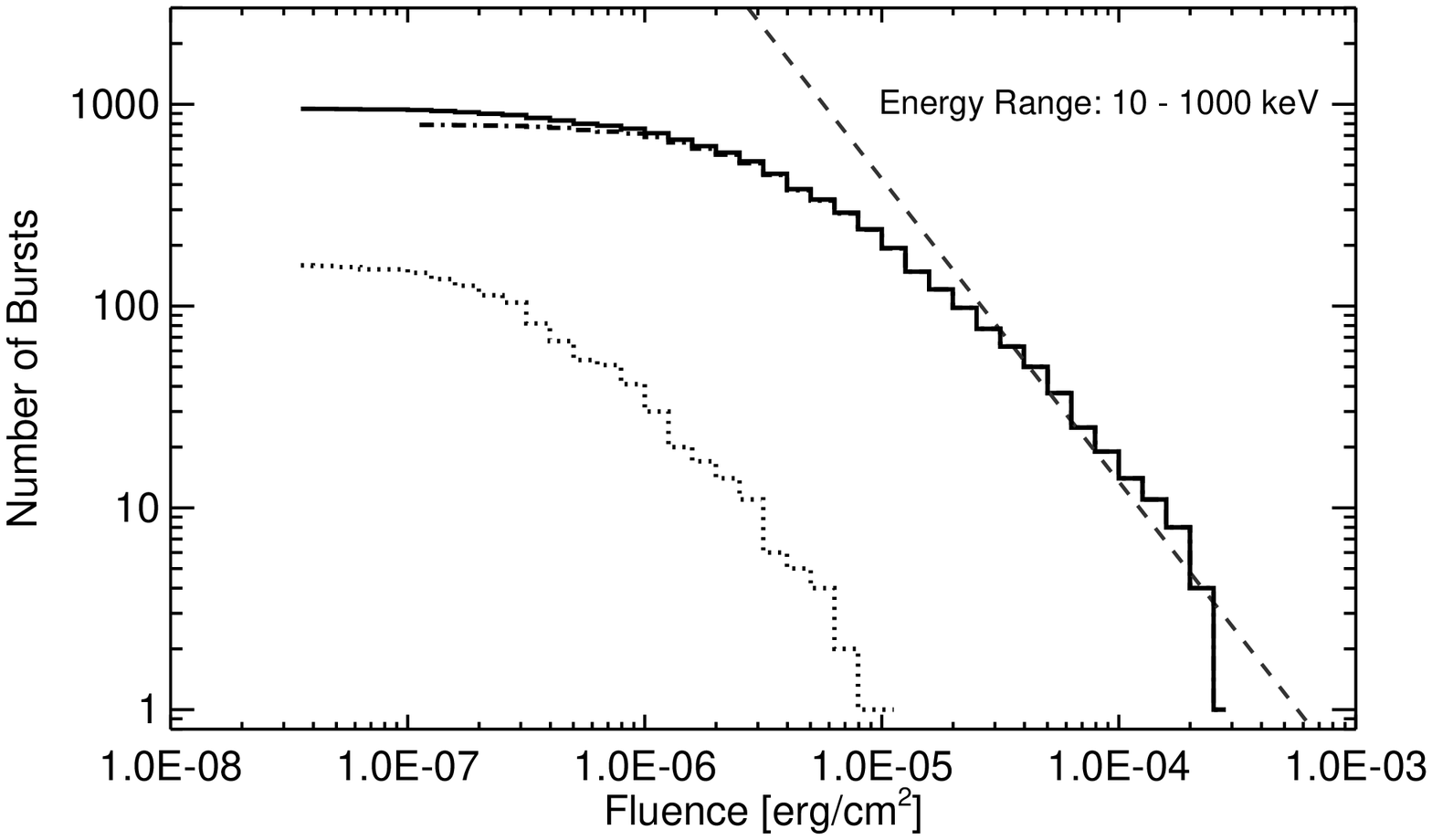}
\plotone{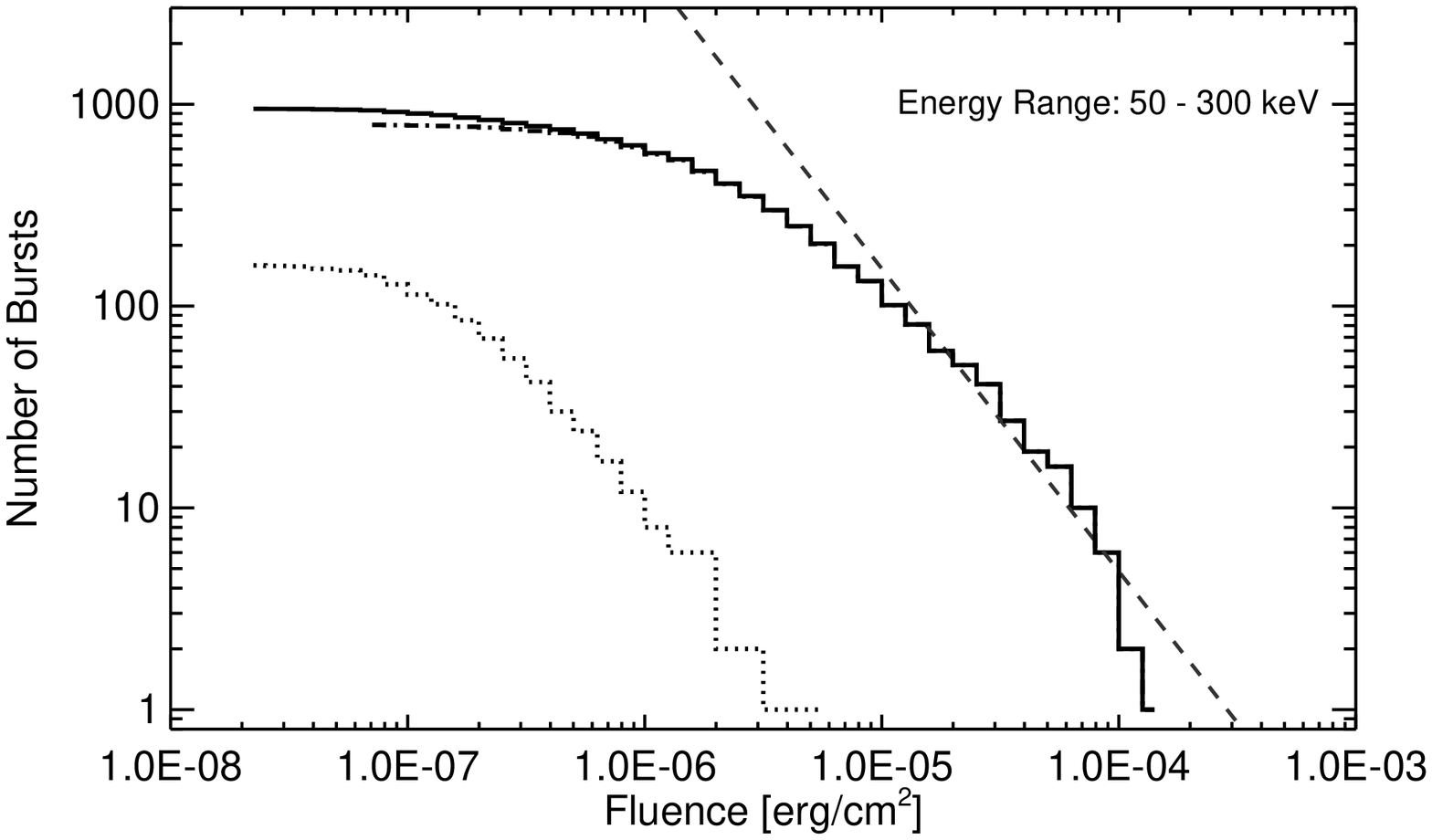}
\caption{\label{flu_fig} Integral distribution of GRB fluence in two energy ranges: 10--1000~keV (upper plot) and 50--300 keV (lower plot). Distributions are shown for the total sample (solid histogram), short GRBs (dots) and long GRBs (dash-dots), using $T_{90} = 2$~s as the distinguishing criterion. In each plot a power law with a slope of $-3 / 2$ (dashed line) is drawn to guide the eye.}
\end{center}
\end{figure}

Integral distributions of the peak fluxes observed for GRBs in the first four years are shown in Figures~\ref{pflx_fig} -- \ref{pf64_fig} for the three different timescales and separately for short and long GRBs. The conclusion made in the first catalog on the shape of the integral distributions is validated. For long GRBs the deviation from the $-3/2$ power-law, expected for spatially homogeneous GRBs, occurs well above the GBM threshold at a flux value of $\sim$10 ph s$^{-1}$ cm$^{-2}$. For  short events the GBM data appear consistent with a homogeneous spatial distribution down to peak flux values around 1~ph s$^{-1}$ cm$^{-2}$ (50 -- 300~keV), below which instrument threshold effects become dominant. The integral fluence distributions for the two energy intervals are shown in Figure~\ref{flu_fig}.

\section{Summary}

The second GBM catalog comprises a list of 953 cosmic GRBs that triggered GBM between 12 July 2008 and 11 July 2012. The now doubled GRB sample establishes the conclusions of the first catalog.
The rate of burst detections per year ($\sim 240$/year), which is only slightly smaller compared to the rate of the BATSE instrument \citep[$\sim 300$/year;][]{1999ApJS..122..465P}, can be explained by GBM's additional range of trigger timescales (primarily the 2~s and 4~s timescales), which are compensating for the higher burst detection threshold of GBM  ($\sim 0.7$ vs. $\sim 0.2$ photons cm$^{-2}$ s$^{-1}$ for BATSE). The distribution of GBM durations is consistent with the well-known bimodality measured previously and the fraction of about 17\% of short GRBs in the GBM sample is somewhat smaller than detected by BATSE, which is attributed mainly to GBM's ability to trigger on longer timescales.

\acknowledgments

Support for the German contribution to GBM was provided by the Bundesministerium f\"ur Bildung und Forschung (BMBF) via the Deutsches Zentrum f\"ur Luft und Raumfahrt (DLR) under contract number 50 QV 0301. A.v.K. was supported by the Bundesministeriums f\"ur Wirtschaft und Technologie (BMWi) through DLR grant 50 OG 1101.
AG acknowledges the support of the Graduate Student Researchers Program funded
by NASA. SMB acknowledges support of the European Union Marie Curie Reintegration
Grant within the 7th Program under contract number PERG04-GA-2008-239176. SF
acknowledges the support of the Irish Research Council for Science, Engineering, and
Technology, co-funded by Marie Curie Actions under FP7. HFY acknowledges support by the DFG cluster of excellence
"Origin and Structure of the Universe".

\clearpage

 \clearpage

\clearpage

\clearpage

\begin{figure}
\begin{center}
\epsscale{1.0}
\plotone{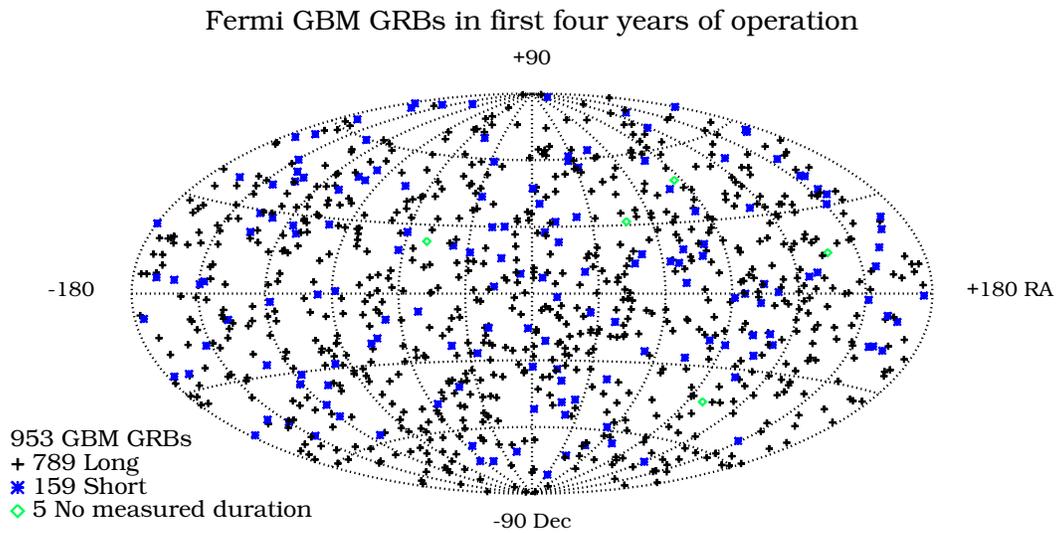}
\caption{\label{sky_dist} Sky distribution of GBM triggered GRBs in celestial coordinates. Crosses indicate long GRBs ($T_{90} > 2$~s); asterisks indicate short GRBs.}
\end{center}
\end{figure}
 \clearpage

 \clearpage

 \clearpage

\clearpage

\clearpage

 \clearpage

\clearpage

\clearpage

\begin{table}[t]
  \centering
  \caption{Trigger statistics of the year 1 \& 2 and year 3 \& 4 catalogs }\label{trigstat1st2ndcat}
\begin{tabular}{|l||c|c|c|c|c|c|c|c|c|}
  \hline
   & GRBs & SGRs & TGFs & SFs & CPs & Other & Sum & ARRs & LAT GRBs \\ \hline
  Year 1 \& 2 & 492\tablenotemark{a} & 170 & 79 & 31 & 69\tablenotemark{b} & 65\tablenotemark{b} & 906\tablenotemark{c} & 40 & 22 \\ \hline
  Year 3 \& 4  & 462 & 17 & 182 & 363 & 138 & 58 & 1220 & 48 & $20\tablenotemark{d}$  \\ \hline
  Year 1 to 4 & 954 & 187 & 261 & 394 & 207 & 123 & 2126 & 88\tablenotemark{e} & 43  \\
  \hline
\end{tabular}
\tablenotetext{a}{The number of GRBs triggers during year 1 \& 2 is including the two triggers on the ultra-long GRB 091024.}
\tablenotetext{b}{ The numbers of non GRB triggers in year 1 \& 2 differ from the numbers cited in \cite{Pacie12}, since some of the triggers were reclassified}
\tablenotetext{c}{The total numbers of triggers is two less compared to \cite{Pacie12}, since the two commanded triggers (bn100709294 \& bn100711145) were not counted.}
\tablenotetext{d}{The three year \Fermi\ LAT GRB catalog \citep{2013ApJS..209...11A} includes bursts only from August 2008 to August 2011 (Year 1 \& 2: 22 GRBs, Year 3 \& 4: 13 GRBs). The seven additional GRB detections from year 3 \&4, are listed in the public GRB list of the \Fermi\ LAT team: \url{http://fermi.gsfc.nasa.gov/ssc/observations/types/grbs/lat_grbs/}}
\tablenotetext{e}{Due to misclassification of events as GRBs by the FSW, some of the ARRs occurred for other event types. There were in addition two positive ARRs for GBM trigger 100701.490 and 110920.546 with no slew, which was disabled at spacecraft level at that time.}
\end{table}




\begin{deluxetable}{ccccccccccccc}

\rotate

\tabletypesize{\small}

\tablewidth{614pt}

\tablecaption{\label{trigger:criteria:history} Trigger Criteria History}


\tablehead{\colhead{Algorithm} & \colhead{Timescale} & \colhead{Offset} & \colhead{Channels} & \colhead{Energy} & \multicolumn{7}{c}{Threshold ($0.1 \sigma$)\tablenotemark{a}}\\
\cline{6-13}\\
\multicolumn{5}{c}{ } & \multicolumn{3}{c}{ 2008} & \multicolumn{4}{c}{2009 } & 2010\\
\colhead{Number} & \colhead{(ms)} & \colhead{(ms)} & \colhead{} & \colhead{(keV)} & \colhead{July 11} & \colhead{July 14} & \colhead{Aug 1} & \colhead{May 8} & \colhead{Oct 29 } & \colhead{Nov 10} & \colhead{Dec 7} & \colhead{Mar 26}}
\startdata
1 & 16 & 0 & 3--4 & 50--300 & 75 & : & : & : & : & : & : & : \\
2 & 32 & 0 & 3--4 & 50--300 & 75 & : & : & : & : & : & : & : \\
3 & 32 & 16 & 3--4 & 50--300 & 75 & : & : & : & : & : & : & : \\
4 & 64 & 0 & 3--4 & 50--300 & 45 & : & 50 & : & : & : & : & : \\
5 & 64 & 32 & 3--4 & 50--300 & 45 & : & 50 & : & : & : & : & : \\
6 & 128 & 0 & 3--4 & 50--300 & 45 & : & 48 & 50 & : & : & : & : \\
7 & 128 & 64 & 3--4 & 50--300 & 45 & : & 48 & 50 & : & : & : & : \\
8 & 256 & 0 & 3--4 & 50--300 & 45 & : & : & : & : & : & : & : \\
9 & 256 & 128 & 3--4 & 50--300 & 45 & : & : & : & : & : & : & : \\
10 & 512 & 0 & 3--4 & 50--300 & 45 & : & : & : & : & : & : & : \\
11 & 512 & 256 & 3--4 & 50--300 & 45 & : & : & : & : & : & : & : \\
12 & 1024 & 0 & 3--4 & 50--300 & 45 & : & : & : & : & : & : & : \\
13 & 1024 & 512 & 3--4 & 50--300 & 45 & : & : & : & : & : & : & : \\
14 & 2048 & 0 & 3--4 & 50--300 & 45 & : & : & : & : & : & : & : \\
15 & 2048 & 1024 & 3--4 & 50--300 & 45 & : & : & : & : & : & : & : \\
16 & 4096 & 0 & 3--4 & 50--300 & 45 & : & : & : & : & : & : & : \\
17 & 4096 & 2048 & 3--4 & 50--300 & 45 & : & : & : & : & : & : & : \\
18 & 8192 & 0 & 3--4 & 50--300 & C & 50 & : & : & D & : & : & : \\
19 & 8192 & 4096 & 3--4 & 50--300 & C & 50 & : & : & D & : & : & : \\
20 & 16384 & 0 & 3--4 & 50--300 & C & 50 & D & : & : & : & : & : \\
21 & 16384 & 8192 & 3--4 & 50--300 & C & 50 & D & : & : & : & : & : \\
22 & 16 & 0 & 2--2 & 25--50 & D & 80 & : & : & : & : & : & : \\
23 & 32 & 0 & 2--2 & 25--50 & D & 80 & : & : & : & : & : & : \\
24 & 32 & 16 & 2--2 & 25--50 & D & 80 & : & : & : & : & : & : \\
25 & 64 & 0 & 2--2 & 25--50 & D & 55 & : & : & : & : & : & : \\
26 & 64 & 32 & 2--2 & 25--50 & D & 55 & : & : & : & : & : & : \\
27 & 128 & 0 & 2--2 & 25--50 & D & 55 & : & : & D & : & : & : \\
28 & 128 & 64 & 2--2 & 25--50 & D & 55 & : & : & D & : & : & : \\
29 & 256 & 0 & 2--2 & 25--50 & D & 55 & : & : & D & : & : & : \\
30 & 256 & 128 & 2--2 & 25--50 & D & 55 & : & : & D & : & : & : \\
31 & 512 & 0 & 2--2 & 25--50 & D & 55 & : & : & D & : & : & : \\
32 & 512 & 256 & 2--2 & 25--50 & D & 55 & : & : & D & : & : & : \\
33 & 1024 & 0 & 2--2 & 25--50 & D & 55 & : & : & D & : & : & : \\
34 & 1024 & 512 & 2--2 & 25--50 & D & 55 & : & : & D & : & : & : \\
35 & 2048 & 0 & 2--2 & 25--50 & D & 55 & : & : & D & : & : & : \\
36 & 2048 & 1024 & 2--2 & 25--50 & D & 55 & : & : & D & : & : & : \\
37 & 4096 & 0 & 2--2 & 25--50 & D & 65 & : & : & D & : & : & : \\
38 & 4096 & 2048 & 2--2 & 25--50 & D & 65 & : & : & D & : & : & : \\
39 & 8192 & 0 & 2--2 & 25--50 & D & 65 & : & : & D & : & :& :  \\
40 & 8192 & 4096 & 2--2 & 25--50 & D & 65 & : & : & D & : & : & : \\
41 & 16384 & 0 & 2--2 & 25--50 & D & 65 & D & : & : & : & : & : \\
42 & 16384 & 8192 & 2--2 & 25--50 & D & 65 & D & : & : & : & : & : \\
43 & 16 & 0 & 5--7 & $> 300$ & D & 80 & : & : & : & : & : & : \\
44 & 32 & 0 & 5--7 & $> 300$ & D & 80 & : & : & D & : & : & : \\
45 & 32 & 16 & 5--7 & $> 300$ & D & 80 & : & : & D & : & : & : \\
46 & 64 & 0 & 5--7 & $> 300$ & D & 55 & : & 60 & D & : & : & : \\
47 & 64 & 32 & 5--7 & $> 300$ & D & 55 & : & 60 & D & : & : & : \\
48 & 128 & 0 & 5--7 & $> 300$ & D & 55 & : & : & D & : & : & : \\
49 & 128 & 64 & 5--7 & $> 300$ & D & 55 & : & : & D & : & : & : \\
50 & 16 & 0 & 4--7 & $> 100$ & D & 80 & : & : & : & : & : & : \\
51 & 32 & 0 & 4--7 & $> 100$ & D & 80 & : & : & D & : & : & : \\
52 & 32 & 16 & 4--7 & $> 100$ & D & 80 & : & : & D & : & : & : \\
53 & 64 & 0 & 4--7 & $> 100$ & D & 55 & : & : & D & : & : & : \\
54 & 64 & 32 & 4--7 & $> 100$ & D & 55 & : & : & D & : & : & : \\
55 & 128 & 0 & 4--7 & $> 100$ & D & 55 & : & : & D & : & : & : \\
56 & 128 & 64 & 4--7 & $> 100$ & D & 55 & : & : & D & : & : & : \\
57 & 256 & 0 & 4--7 & $> 100$ & D & 55 & : & : & D & : & : & : \\
58 & 256 & 128 & 4--7 & $> 100$ & D & 55 & : & : & D & : & : & : \\
59 & 512 & 0 & 4--7 & $> 100$ & D & 55 & : & : & D & : & : & : \\
60 & 512 & 256 & 4--7 & $> 100$ & D & 55 & : & : & D& :  & : & : \\
61 & 1024 & 0 & 4--7 & $> 100$ & D & 55 & : & : & D & : & : & : \\
62 & 1024 & 512 & 4--7 & $> 100$ & D & 55 & : & : & D & : & : & : \\
63 & 2048 & 0 & 4--7 & $> 100$ & D & 55 & : & : & D & : & : & : \\
64 & 2048 & 1024 & 4--7 & $> 100$ & D & 55 & : & : & D & : & : & : \\
65 & 4096 & 0 & 4--7 & $> 100$ & D & 65 & : & : & D & : & : & : \\
66 & 4096 & 2048 & 4--7 & $> 100$ & D & 65 & : & : & D & : & : & : \\
&   &  & 5--7 & $> 300$  &  &  &  &  &  & 60 & 55 & : \\
\rb{116\tablenotemark{b}} & \rb{16} & \rb{0} & BGO/3--6 & 2 - 40 MeV & \rb{D} & \rb{:} & \rb{:} & \rb{:} & \rb{:} & 55 & 45 & : \\
 &   &  & 5--7 & $> 300$ &  &  &  &  & & 55 & 45 & : \\
\rb{117\tablenotemark{b}} & \rb{16} & \rb{0} & BGO/3--6 & 2 - 40 MeV & \rb{D} & \rb{:} & \rb{:} & \rb{:} & \rb{:} & 55 & 45 & : \\
 &   &  & 5--7 & $> 300$ &  &  &  &  & & 55 & 45 & : \\
\rb{118\tablenotemark{b}} & \rb{16} & \rb{0} & BGO/3--6 &  2 - 40 MeV & \rb{D} & \rb{:} & \rb{:} & \rb{:} & \rb{:} & 55 & 45 & : \\
\raisebox{0.0ex}{119\tablenotemark{b}} & 16 & 0 & BGO/3--6 &  2 - 40 MeV  & D & : & : & : & : & 55 & 45 & 47\\
\enddata
\tablenotetext{a}{Symbol ':' indicates no change from previous setting; 'C' indicates that the algorithm is in compute mode (see text); 'D' indicates that the algorithm is disabled.}
\tablenotetext{b}{Trigger algorithms using the BGO detector count rates. Algorithm 116 triggers off when at least two NaI and  one BGO detectors are exceeding the trigger threshold.  Algorithms 117 is same as 116, but impose the additional requirement that the triggered detectors are on the +X side of the spacecraft.  Algorithm 118 is the same as 117, but requiring the triggered detectors to be on the -X side of the spacecraft. Algorithm 119 requires a significant rate increase in both BGO detectors.}


\end{deluxetable}

\begin{table}[t]
  \centering
  \caption{GBM GCN notice types (For more details see: \url{http://gcn.gsfc.nasa.gov/fermi.html\#tc2)}}\label{GCNnoticetypes}
\begin{tabular}{|l||p{2cm}|p{7cm}|c|}
  \hline
GCN/FERMI & Sequence of Notices & Content / Purpose & Issues\\ \hline
\_GBM\_ALERT & 1$^{\rm st}$, occurs directly after GBM trigger  & date, time, trigger criteria, trigger detection significance, algorithm used to make the detection & 1 \\ \hline
\_GBM\_FLT\_POSITION & 2$^{\rm nd}$ & RA, DEC GRB location, calculated by on-board flight software  & 1 - 5 \\ \hline
\_GBM\_GND\_POSITION & 3$^{\rm rd}$ &  RA, DEC GRB location, calculated by automated ground software  &  $\ge 0$ \\ \hline
\_GBM\_FINAL\_POSITION & 4$^{\rm th}$ & H-i-t-l location. If the trigger is a GRB and it is not detected by an instrument with better location accuracy, a GBM final notice is sent within 2 hours & $\ge 0$ \\ \hline
\_SC\_SLEW & only in case of an ARR &  indicates whether or not the spacecraft determined if it will slew to this burst. $\sim 1 - 3$/month & 1 \\ \hline
\end{tabular}
\end{table}

\begin{table}[t]
  \centering
  \caption{Trigger algorithm statistics}\label{trigstatalgor}
\begin{tabular}{|c|c|c|c|c|c|c|c|c|c|}
  \hline
  Algorithm & Time ms & Energy keV & GRBs & SGRs & TGFs & SFs & CPs & Other & Comment\tablenotemark{a} \\ \hline
 1 - 5 & 16 -64  & 50 - 300        & 163 & 72 & 5   &  1  & 8   & 10 & GRB \\ \hline
 6 - 11 & 128 - 512  &  50 - 300  & 351 & 7  & -   & 6  & 6  & 31 & GRB \\ \hline
  12 - 17 & 1024 - 4096  &  50 - 300  & 418 & -  & -   & 35  & 150  & 25 & GRB\\ \hline
  18 - 21 & 8192 - 16384 &  50 - 300 & 3 & -  & -   & -  & - & - & D\\ \hline
  22 - 26 & 16 -64 & 25 - 50        & 7   & 105 & -   & 349 & 7   &  5 & SGR\\ \hline
  27 - 32 & 128 - 512  & 25 - 50   & 2  & 3  & -   &  1  & 2   &  - & D\\ \hline
  33 - 38 & 1024 - 4096 & 25 - 50   & 8   & -  & -   &  2  & 11   &  3 & D\\ \hline
  39 - 42 &  8192 - 16384 & 25 - 50   & 1   & -  & -   &  -  & 8   &  3 & D\\ \hline
    43  & 16 &$> 300$         & -   & -  & 30  &  -  & -   &  1 & TGF\\ \hline
  44 - 49 & 32 - 128&$> 300$         & -   & -  & -  &  -  & -   &  5 & D\\ \hline
  50  & 16  & $> 100$      & -   & -  & 5   &  -  & 4   &  - & TGF\\ \hline
  51 - 66 & 32 - 4096 & $> 100$      & 1   & -  &  -  &  -  &  -   &  - & D\\ \hline
 116 - 119 & 16 & BGO               & -   & -  & 221 &  -  & 11   &  40 & TGF\\ \hline
\end{tabular}
\tablenotetext{a}{'GRB', 'SGR' and 'TGF' indicate the source classes that are most likely to trigger the corresponding algorithm; 'D' indicates that the algorithm was finally disabled at the end of year 4.}
\end{table}

\begin{table}[t]
  \centering
  \caption{Breakdown of long and short GRBs which triggered on BATSE- and non-BATSE-like GBM GRB trigger algorithms, individually listed for the year 1 \& 2, year 3 \& 4  and full four year catalog periods. The fraction of short GRBs (in \%) with respect to the total number of observed GRBs is  stated for all GBM GRBs and GRBs which have triggered on BATSE-like trigger algorithms.}\label{grbbreakdown}
\begin{tabular}{|l||c|c|c|c|c|}
  \hline
 &  & GRBs & long &  short GRBs  & no  \\
\rb{Years} & \rb{Algorithm} & ($1^{\rm st}$ / $2^{\rm nd}$ half-bin\tablenotemark{a}) & GRBs & (Range\tablenotemark{b}) & duration \\ \hline
 & & & & 88 (18\%) &  \\
 & \rb{ALL} & \rb{491} & \rb{400\tablenotemark{c}} & (73 (15\%) - 104 (21\%)) & \rb{3\tablenotemark{c}} \\ \cline{2-5}
 & & 419 & & & \\
\rb{1 \& 2} & \rb{BATSE}  & (405 / 408) & \rb{336} & \rb{83 (20\%)} & \\ \cline{2-5}
 & non-BATSE & 68 & 63 & 5 &  \\ \hline
  & & & & 71 (15\%) &  \\
 & \rb{ALL} & \rb{462} & \rb{389} & (51 (11\%) - 89 (19\%)) & \rb{2} \\ \cline{2-6}
 & & 395 & & & \\
\rb{3 \& 4} & \rb{BATSE}  & (372 / 366) & \rb{330} & \rb{63 (16\%)} & \rb{2} \\ \cline{2-6}
 & non-BATSE & 67 & 59 & 8 & - \\ \hline
  & & & & 159 (17\%) &  \\
 & \rb{ALL} & \rb{953} & \rb{789} & (124 (13\%) - 193 (20\%)) & \rb{5} \\ \cline{2-6}
 & & 814 & & &\\
\rb{1 to 4} & \rb{BATSE}  & (777 / 774) & \rb{666} & \rb{146 (18\%)} & \rb{2}\\ \cline{2-6}
 & non-BATSE & 135 & 122 & 13 &  - \\ \hline
\end{tabular}
\tablenotetext{a}{Number of GRBs which triggered on BATSE-like search algorithms which are offset half the timescale ($2^{\rm nd}$ half-bin) compared to the original BATSE-like algorithms ($1^{\rm st}$ half-bin).}
\tablenotetext{b}{Number of short GRBs within the quoted duration errors (see Table \ref{durations})}
\tablenotetext{c}{The ultra-long GRB 091024A which triggered GBM twice and the three GRBs with no measured duration weren't considered for the BATSE/non-BATSE classification of the years 1 \& 2 triggers.}
\end{table}




 commands
\tablenotetext{a}{Bursts with Trigger ID and GRB Name in italics have significant emission in at least one BGO detector (see text).}
\tablenotetext{b}{Other instrument detections: Mo: Mars Observer , K: Konus-Wind, R: RHESSI, IA: \INTEGRAL\ SPI-ACS, IS: \INTEGRAL\ IBIS-ISGRI, S: \Swift, Me: Messenger, W: \Suzaku, A: \AGILE, M: \MAXI, L: \Fermi\ LAT}
\tablenotetext{c}{GRB091024A triggered GBM twice.}



\end{deluxetable}




 commands
\tablenotetext{a}{Data problems precluded duration analysis.}
\tablenotetext{b}{Used TTE binned at 32 ms.}
\tablenotetext{c}{Partial earth occultation is likely; durations are lower limits.}
\tablenotetext{d}{Possible precursor at $\sim T_0-120$ s.}
\tablenotetext{e}{Data cut off due to SAA entry while burst in progress; durations are lower limits.}
\tablenotetext{f}{SAA entry at $T_0+83$~s; durations are lower limits.}
\tablenotetext{g}{Used TTE binned at 16 ms.}
\tablenotetext{h}{GRB091024 triggered GBM twice.}
\tablenotetext{i}{Too weak to measure durations; visual duration is $\sim 0.025$~s.}
\tablenotetext{j}{Possible contamination due to emergence of Crab \& A0535+26 from Earth occultation.}
\tablenotetext{k}{Solar activity starting at $T_0+200$~s. Post burst background interval was selected  before.}
\tablenotetext{l}{Data cut off due to SAA entry while burst in progress;it is not possible to determine durations.}
\tablenotetext{m}{Spacecraft in sun pointing mode, detector threshold raised, location of burst nearly in -z direction. The response, peak fluxes and fluence in the $10 -100$~keV energy range have large errors. Fluence, peak fluxes and durations in BATSE energy range (50 -300 keV) are reliable.}
\tablenotetext{n}{{L}ocalization of precursor at $T_0-120$~s is consistent with burst location and was included in the duration analysis.}
\tablenotetext{o}{SAA entry at $T_0+100$~s; durations are lower limits.}
\tablenotetext{p}{TTE/CTTE data not available, 64~ms peak fluxes may not be correct.}



\end{deluxetable}




 commands



\end{deluxetable}

\end{document}